\documentclass[conference]{IEEEtran}
\IEEEoverridecommandlockouts

\usepackage[utf8]{inputenc}
\usepackage{cite}
\usepackage{amsmath,amssymb,amsfonts}
\usepackage{algorithmic}
\usepackage{graphicx}
\usepackage{textcomp}
\usepackage{xcolor}
\usepackage{fancyhdr}
\usepackage[hyphens]{url}
\usepackage{multicol}
\usepackage{subfigure}
\usepackage{pifont}
\usepackage{amsmath,amsfonts,amssymb}
\usepackage{enumitem}
\usepackage{multirow}
\usepackage[export]{adjustbox}
\usepackage{picins}
\usepackage{cleveref}
\crefformat{section}{\S#2#1#3}
\crefformat{subsection}{\S#2#1#3}
\crefformat{subsubsection}{\S#2#1#3}

\usepackage{titlesec}
\titlespacing\section{3pt}{3pt plus 2pt minus 2pt}{1pt plus 1pt minus 1pt}
\titlespacing\subsection{3pt}{2pt plus 2pt minus 1pt}{2pt plus 1pt minus 1pt}
\titlespacing\subsubsection{3pt}{2pt plus 2pt minus 1pt}{2pt plus 1pt minus 1pt}

\def\BibTeX{{\rm B\kern-.05em{\sc i\kern-.025em b}\kern-.08em
    T\kern-.1667em\lower.7ex\hbox{E}\kern-.125emX}}

\begin{document}

\title{\LARGE \bf EXMA: A Genomics Accelerator for Exact-Matching\thanks{This work was partially supported by the National Science Foundation (NSF) through awards CCF-1908992 and CCF-1909509.}} 

\author{
\IEEEauthorblockN{Lei Jiang}
\IEEEauthorblockA{\textit{Indiana University Bloomington} \\
\texttt{jiang60@iu.edu}}
\and
\IEEEauthorblockN{Farzaneh Zokaee}
\IEEEauthorblockA{\textit{Indiana University Bloomington} \\
\texttt{fzokaee@iu.edu}}
}

\maketitle

\begin{abstract}
Genomics is the foundation of precision medicine, global food security and virus surveillance. Exact-match is one of the most essential operations widely used in almost every step of genomics such as alignment, assembly, annotation, and compression. Modern genomics adopts Ferragina-Manzini Index (FM-Index) augmenting space-efficient Burrows-Wheeler transform (BWT) with additional data structures to permit ultra-fast exact-match operations. However, FM-Index is notorious for its poor spatial locality and random memory access pattern. Prior works create GPU-, FPGA-, ASIC- and even process-in-memory (PIM)-based accelerators to boost FM-Index search throughput. Though they achieve the state-of-the-art FM-Index search throughput, the same as all prior conventional accelerators, FM-Index PIMs process only one DNA symbol after each DRAM row activation, thereby suffering from poor memory bandwidth utilization.

In this paper, we propose a hardware accelerator, EXMA, to enhance FM-Index search throughput. We first create a novel EXMA table with a multi-task-learning (MTL)-based index to process multiple DNA symbols with each DRAM row activation. We then build an accelerator to search over an EXMA table. We propose 2-stage scheduling to increase the cache hit rate of our accelerator. We introduce dynamic page policy to improve the row buffer hit rate of DRAM main memory. We also present CHAIN compression to reduce the data structure size of EXMA tables. Compared to state-of-the-art FM-Index PIMs, EXMA improves search throughput by $4.9\times$, and enhances search throughput per Watt by $4.8\times$.
\end{abstract}

\begin{IEEEkeywords}
Domain-Specific Hardware Accelerator, Genomics, Exact-Matching
\end{IEEEkeywords}

\section{Introduction}
\label{s:intro}

Because of the huge advancement of sequencing technologies such as Illumina~\cite{Schirmer:NAR2015}, PacBio SMRT~\cite{Mosher:JMM2014}, and Oxford Nanopore~\cite{Eisenstein:Oxford2012}, sequencing a entire human genome requires only $<1$ day. The big genomic data has been a cornerstone to enabling personalized healthcare~\cite{Merker:Nature2018}, and ensuring global food security~\cite{Ma:TBIO2017}. Recently, genome sequencing becomes a powerful tool to fight virus outbreaks, e.g., Ebola~\cite{Hoenen:EID2016}, Zika~\cite{Quick:NP2017} and COVID-19~\cite{Zhu:NEJM2020}.

However, it is challenging to process and analyze huge volumes of genomic data generated by high throughput sequencers that scale faster than Moore's Law~\cite{Canzar:IEEE2017}. For instance, thousands of USB-drive-size Oxford Nanopore Minion sequencers are deployed to monitor virus outbreaks~\cite{Hoenen:EID2016,Quick:NP2017,Zhu:NEJM2020} in the wild by generating several terabytes data per day. Analyzing a single genome may take hundreds of CPU hours~\cite{Turakhia:ASPLOS2018,Fuijiki:ISCA2018} even on high-end servers. To overcome the looming crisis of big genomic data, the application-specific hardware acceleration has become essential for genomics.

A genome sequencing pipeline~\cite{Merker:Nature2018} sequences organic genomes, archives genomic data, analyzes genome sequences, and generates genetic variants that can used for patient treatment. Therefore, the \textit{latency} of genome sequencing is a matter of life and death. \textit{Read alignment}~\cite{Li:BWAMEM2013}, which aligns reads, i.e., small DNA fragments, against a long genome reference, is identified as one of the most time-consuming steps~\cite{Chang:FCCM2016,Turakhia:ASPLOS2018,Fuijiki:ISCA2018,Zokaee:PACT2019,Huangfu:MICRO2019} in genome analysis. Read alignment adopts the \textit{seed-and-extend} paradigm~\cite{Li:BWAMEM2013,Schmidt:NC2019}, and thus includes two major stages, i.e., \textit{seeding} and \textit{seed extension}. During seeding, parts of each read are mapped to their exactly matched positions, i.e., seeds, of the long reference by hash tables~\cite{Turakhia:ASPLOS2018,Fuijiki:ISCA2018,Nag:MICRO2019} or Ferragina-Manzini Index (FM-Index)~\cite{Schmidt:NC2019,Li:BWAMEM2013}. Seed extension pieces together a larger sequence with seeds and edit distance errors, i.e., insertions, deletions (indels) and substitutions, by dynamic programming~\cite{Kaplan:MICRO2017,Madhavan:ISCA2014,Enzo:ICBB2017}.

\begin{figure}[t!]
\centering
\includegraphics[width=3.3in]{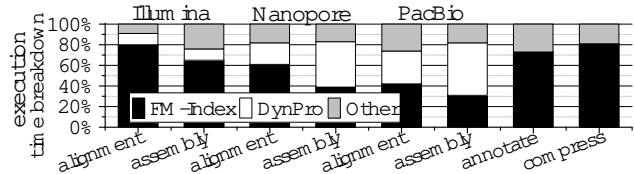}
\vspace{-0.1in}
\caption{Execution time breakdown of human genome analysis (\texttt{DynPro} means dynamic programming).}
\label{f:dna_moti_fig}
\vspace{-0.25in}
\end{figure}

State-of-the-art read alignment applications such as BWA-MEM~\cite{Li:BWAMEM2013}, MA~\cite{Schmidt:NC2019} and SOAP~\cite{Luo:PLOS2013} use FM-Index to build super-maximal exact matches (SMEMs) during seeding, since it augments the space-efficient Burrows-Wheeler transform (BWT)~\cite{Burrows:HSRR1994} with accessory data structures that permit ultra-fast exact-match operations. SMEMs generated by FM-Index guarantee each seed does not overlap other seeds and has the maximal length that cannot be further extended. Compared to hash tables, FM-Index reduces not only the number of errors in output genome mappings but also the durations of seed extension substantially~\cite{Ahmed:ICBB2016}.

Besides read alignment, FM-Index is widely used for exact-match operations in other time-consuming steps of genome analysis such as genome assembly~\cite{Simpson:GR2012}, annotation~\cite{Healy:GEN2003} and compression~\cite{Prochazka:DCC2014}. Figure~\ref{f:dna_moti_fig} shows the execution time breakdown of various genome analysis applications on a human genome\footnote{The experimental methodology is elaborated in~\cref{s:eandm}.}. On average, \textbf{FM-Index searches cost} $\mathbf{31\%}$ $\sim$ $\mathbf{81\%}$ \textbf{of the execution time of these genome analysis applications}. Since Illumina, Nanopore and PacBio genome sequencers generate reads having different lengths and error rates, aligning and assembling these reads require different amounts of time for FM-Index searches. The reads produced by Illumina machines have lower error rates, so an Illumina dataset invokes FM-Index searches more frequently.

However, FM-Index is notorious for its poor spatial locality and random memory access pattern~\cite{Chacon:TCBB2015}. The kernel of conventional FM-Index search is pointer chasing. After activating one DRAM row, FM-Index processes only one DNA symbol, thereby greatly decreasing DRAM bandwidth utilization. Although a recent algorithmic work, LISA~\cite{Ho:WSMN2019}, uses a learned index~\cite{Kraska:ICMD2018} to search multiple DNA symbols after each row activation, the learned index accuracy is low. LISA has to search many unnecessary entries, and thus achieves only moderate search throughput improvement. Beyond CPUs~\cite{Li:BWAMEM2013} and GPUs~\cite{Luo:PLOS2013}, prior work creates FPGA~\cite{Chang:FCCM2016,Arram:TCBB2017}-, ASIC~\cite{Wu:ITBCS2017}-, and even processing-in-memory (PIM)~\cite{Huangfu:MICRO2019,Zokaee:PACT2019}-based designs to accelerate conventional FM-Index searches processing only one DNA symbol after each row activation. Therefore, these accelerators are fundamentally limited by the poor off-chip memory bandwidth utilization. Instead of searching multiple DNA symbols after each row activation, a recent DRAM PIM, MEDAL~\cite{Huangfu:MICRO2019}, achieves the state-of-the-art search throughput by enabling DRAM chip-level parallelism, where each chip can independently activate a partial row to process a DNA symbol. However, we observe that there are a lot of conflicts on the DDR4 address bus shared by all chips in a rank. The shared address bus seriously limits search throughput of MEDAL.

In this paper, we propose an algorithm and hardware co-designed accelerator, \textbf{EXMA}, to process \textbf{EX}act-\textbf{MA}tch operations during genome analysis. Our contributions are summarized as follows:

\begin{itemize}[nosep,leftmargin=*]
\item \textbf{An EXMA table with a MTL-based index --} We propose a novel data structure, EXMA table, that can process $k$ DNA symbols, i.e., a $k$-mer, in a DRAM row in each FM-Index search iteration. We further present a multi-task-learning (MTL)-based index to accelerate searches over an EXMA table. The MTL-based index trained with multiple $k$-mers uses less neural network parameters, but obtains higher accuracy over learning to search each $k$-mer independently.

\item \textbf{A hardware accelerator --} We build an accelerator to search an EXMA table with a MTL-based index. We present a 2-stage scheduling to increase the hit rate of on-chip caches of our accelerator for the table and its index. We also propose dynamic page policy to improve the row buffer hit rate of DRAM main memory. At last, we introduce CHAIN compression to greatly reduce the data structure size of an EXMA table.

\item \textbf{Search throughput and throughput per Watt --} We evaluated and compare EXMA to prior CPU-, GPU-, FPGA-, ASIC-, and PIM-based FM-Index accelerators. Compared to the state-of-the-art DRAM PIM MEDAL, EXMA improves search throughput by $4.9\times$, and enhances search throughput per Watt by $4.8\times$.
\end{itemize}

\section{Background}
\label{s:back}

\subsection{Read Alignment}

\begin{figure}[t!]
\centering
\subfigure[Read alignment]{
   \includegraphics[width=2in]{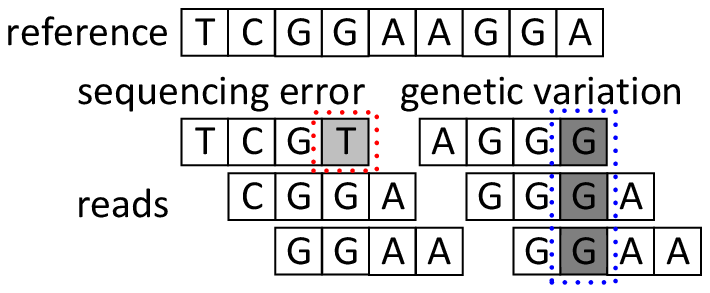}
   \label{f:dna_read_align}
}
\hspace{-0.1in}
\subfigure[seed-\&-extend]{
   \includegraphics[width=1.1in]{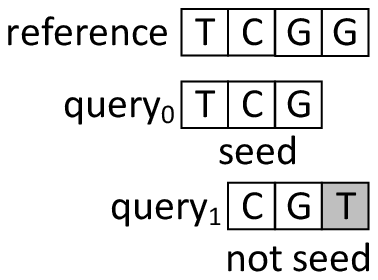}
   \label{f:dna_seed_ex}
}
\vspace{-0.1in}
\caption{Read alignment}
\label{f:dna_seed_extension}
\vspace{-0.3in}
\end{figure}

\textbf{Seed-and-Extend}: As one of the bottlenecks in genome analysis, read alignment may consume hundreds of CPU hours~\cite{Fuijiki:ISCA2018,Chang:FCCM2016,Turakhia:ASPLOS2018}. During read alignment, DNA reads generated by various sequencing machines, e.g., Illumina, PacBio SMRT, and Oxford Nanopore, are mapped to a pre-existing genome reference, as shown in Figure~\ref{f:dna_read_align}. Read alignment is complicated by the fact that there are genetic variations in the human population, and sequencing machines also introduce sequencing errors~\cite{Quail:BMC2012}. The overall variation of human population has been estimated as $0.1\%$~\cite{Canzar:IEEE2017}, while the sequencing error rate of various sequencing machines is $0.2\%\sim30\%$~\cite{Quail:BMC2012}. To reduce sequencing errors, a sequencing machine produces $30\sim50$ reads to cover every position in the genome. As Figure~\ref{f:dna_seed_ex} shows, read alignment adopts the seed-and-extend paradigm~\cite{Wang:BMC2018,Huang:BMC2017,Simpson:GR2012,Li:BWAMEM2013,Li:BIOINFO2012,Schmidt:NC2019} to accommodate sequencing errors and genetic variations. During seeding, a read is divided into multiple smaller parts that are aligned against the reference. If a part is exactly matched, it becomes a seed. The computationally expensive Smith-Waterman algorithm~\cite{Fuijiki:ISCA2018,Chang:FCCM2016,Turakhia:ASPLOS2018} is invoked only around seeds to handle sequencing errors and genetic variations. 

\textbf{Exact-Match Operation}: The alphabet $\sum$ of DNA includes $A$, $C$, $G$ and $T$. Given a genome reference $\mathcal{G}\in\sum^*$ of length $|\mathcal{G}|$ and a query $Q\in\sum^*$ of length $|Q|$, the seeding, aka exact-match problem, is to find all occurrences of $Q$ in $\mathcal{G}$. A na\"{\i}ve algorithm of exhausting all possible positions for $Q$ will take $\mathcal{O}(|\mathcal{G}||Q|)$ comparisons, which is infeasible for large genome. It is possible to use a hash table~\cite{Turakhia:ASPLOS2018,Fuijiki:ISCA2018,Nag:MICRO2019} to support exact-match operations with hundreds of gigabytes DRAM, but the hash-table-based seeding not only degrades genome mapping quality but also prolongs seed extension durations~\cite{Ahmed:ICBB2016}. State-of-the-art alignment algorithms~\cite{Li:BWAMEM2013,Schmidt:NC2019,Luo:PLOS2013} use FM-Index for seeding. To search a query $Q$ over the reference genome $\mathcal{G}$, FM-Index occupies $\mathcal{O}(|\mathcal{G}|log(|\mathcal{G}|))$ DRAM space and does $\mathcal{O}(|Q|)$ comparisons during a search.

\begin{figure*}[htbp!]
\centering
\includegraphics[width=6.2in]{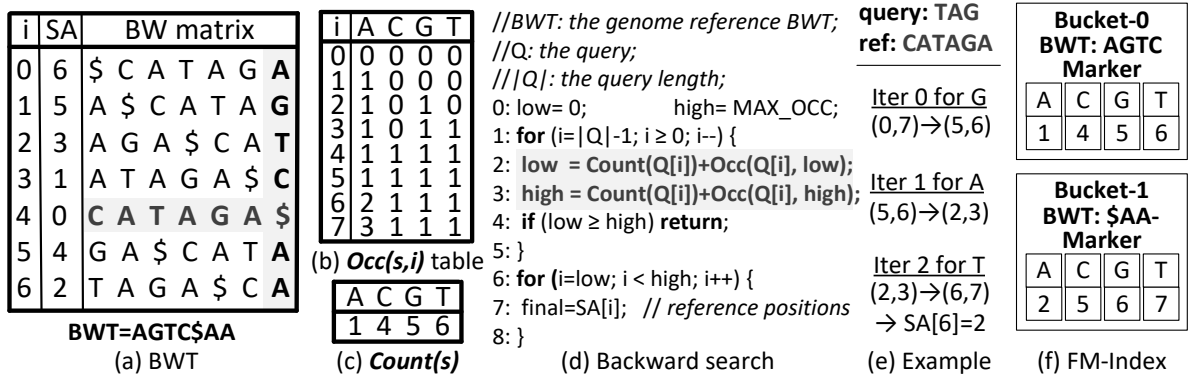}
\vspace{-0.1in}
\caption{FM-Index overview: (a) the BWT of a genome reference $\mathcal{G}=CATAGA\$$; (b) an $Occ(s,i)$ table; (c) a $Count(s)$ table; (d) backward search; (e) a search example; and (f) a bucket-based data structure}
\label{f:dna_fmindex_basic}
\vspace{-0.3in}
\end{figure*}

\subsection{FM-Index}

\subsubsection{Burrows-Wheeler Transform}

FM-index is built upon BWT~\cite{Burrows:HSRR1994}. To compute the BWT of a genome reference $\mathcal{G}$, we can list all its circularly shifted sequences. For instance, Figure~\ref{f:dna_fmindex_basic}(a) shows $\mathcal{G}=CATAGA\$$, where $\$$ indicates the end of the sequence and it is the lexicographically smallest symbol. Circularly shifted sequences of $\mathcal{G}=CATAGA\$$ can be listed as $\$CATAGA$, $A\$CATAG$, $\ldots$, and $CATAGA\$$, which can be sorted in the \textit{lexicographical} order to form a Burrows-Wheeler (BW)-matrix. The last column of the BW-matrix is the BWT of $\mathcal{G}$, i.e., $BWT(\mathcal{G})=AGTC\$AA$. The sub-sequence ending with $\$$ in each row of the BW-matrix is a suffix of $\mathcal{G}$, which can be denoted by an integer (SA in Figure~\ref{f:dna_fmindex_basic}(a)) recording its starting position in the reference. For example, $ATAGA\$$ is $SA[3]=1$, which means it starts from the position 1 of $\mathcal{G}$.

\subsubsection{FM-index}
The data structure and search algorithm of FM-Index can be summarized as:
\begin{itemize}[nosep,leftmargin=*]
\item \textbf{Occ and Count}. FM-index searches are implemented with two functions $Occ(s,i)$ and $Count(s)$ over the BWT of $\mathcal{G}$. As Figure~\ref{f:dna_fmindex_basic}(b) exhibits, $Occ(s,i)$ returns the number of symbol $s$ in the BWT from the position $0$ to the position $i-1$, e.g., $Occ(C,5)=1$, which means that there is only 1 $C$ from the position 0 to the position 4 of $BWT(\mathcal{G})=AGTC\$AA$. $Count(s)$ shown in Figure~\ref{f:dna_fmindex_basic}(c) computes the number of symbols in the BWT that are lexicographically smaller than the symbol $s$, e.g., $Count(T)=6$, which indicates that there are 6 symbols in $BWT(\mathcal{G})=AGTC\$AA$ lexicographically smaller than $T$. 

\item \textbf{Backward Search}. An exact-match operation is implemented by backward search, whose algorithm can be viewed in Figure~\ref{f:dna_fmindex_basic}(d). The interval $(low, high)$ covers a range of indices in the BW-matrix where the suffixes have the same prefix. The pointer $low$ locates the index in the BW-matrix where the pattern is first found as a prefix, while the pointer $high$ provides the index after the one where the pattern is last found. At first, $low$ and $high$ are initialized to the minimum and maximum indexes of the $Occ$ table respectively. And then, they iterate from the \textit{last} symbol in a query $Q$ to the first. The pointer $pos$ is updated by
\vspace{-0.05in}
\begin{equation}
\centering
Count(Q[i])+Occ(Q[i],pos)
\label{e:dna_fm_lpm}
\vspace{-0.05in}
\end{equation}
where $Q[i]$ indicates the $i_{th}$ symbol in the query $Q$. The pointer $pos$ can be $low$ or $high$, as shown from the lines 2 to 3 in Figure~\ref{f:dna_fmindex_basic}(d). The computations of $low$ and $high$ are \textbf{pointer chasing} and thus suffer from poor spatial locality~\cite{Arram:TCBB2017,Zokaee:PACT2019,Huangfu:MICRO2019}. Finally, the interval $(low, high)$ gives the range of indexes in the BW-matrix where the suffixes have the target query as a prefix. These indexes are converted to reference genome positions using SA. Figure~\ref{f:dna_fmindex_basic}(e) illustrates an example of searching a query $TAG$ in the reference $\mathcal{G}=CATAGA\$$. Before a search happens, $(low, high)$ is initialized to $(0,7)$. In the iteration 0, the last symbol $G$ is processed, and then $(low, high)$ is updated to $(5,6)$. After three iterations, $(low, high)$ eventually equals $(6,7)$. By looking up $SA[6]=2$ in Figure~\ref{f:dna_fmindex_basic}(a), we find that the query $TAG$ in reference $\mathcal{G}=CATAGA\$$ starts from the position 2. 

\item \textbf{Bucket-based Storage}. Both $Count(s)$ and $Occ(s,i)$ can be pre-calculated and stored. However, the storage overhead of $Occ(s,i)$ is proportional to the genome reference length $|\mathcal{G}|$, and thus significant. To keep the storage overhead in check, the $Occ(s,i)$ values are sampled into buckets of width $d$ shown in Figure~\ref{f:dna_fmindex_basic}(f) ($d=4$). The $Occ(s,i)$ values are stored each $d$ positions as markers to reduce the storage overhead by a factor of $d$. The omitted $Occ(s,i)$ values can be reconstructed by summing the previous marker and the number of symbol $s$ from the remaining positions in the BWT bucket. To simplify searches, $Count(s)$ values of each symbol are added to corresponding markers. Markers and BWT buckets are interleaved to build a FM-Index. 
\end{itemize}

\subsubsection{Multi-step FM-Index}
\vspace{-0.1in}
\pichskip{5pt}
\parpic[l][b]{
\begin{minipage}{1in}
\centering
\includegraphics[width=1in]{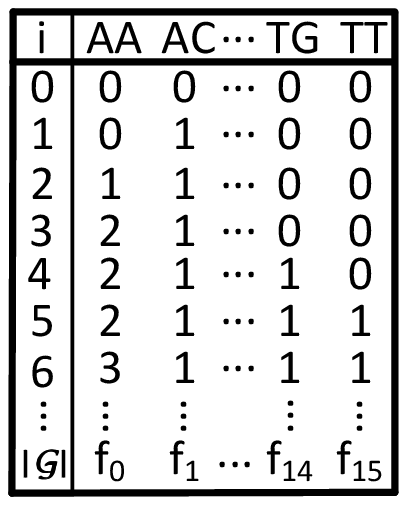}
\vspace{-0.25in}
\caption{A 2-step table.}
\vspace{0.02in}
\label{f:dna_occ_mult88}
\end{minipage}
}
During an iteration of the FM-Index backward search, two memory accesses for $low$ and $high$ are issued for each symbol in a query $Q$. Totally, $2|Q|$ memory accesses are required for an exact match operation of a query. The FM-Index backward search performance is seriously limited by random memory accesses~\cite{Arram:TCBB2017,Zokaee:PACT2019,Huangfu:MICRO2019}, since each access opens a DRAM row but fetches only $64B$. $k$-step FM-Index~\cite{Chacon:PCS2013} is proposed to reduce the number of memory accesses to $\frac{2|Q|}{k}$ by updating a $k$-mer, i.e., $k$ DNA symbols, in each search iteration. The idea of $k$-step FM-Index is to enlarge the alphabet size from $\Sigma$ to $\Sigma^k$. For instance, if $k=2$, instead of single DNA symbols, as Figure~\ref{f:dna_occ_mult88} shows, the enlarged alphabet includes 16 2-mers: $AA$, $AC$, $\ldots$, $TT$. We can construct a BWT with the enlarged alphabet and its corresponding FM-Index to perform $k$-step backward search in the same way. The trade-off for $k$-step FM-Index is the increase in its size, which is calculated as 
\begin{equation}
\centering
F=\frac{\lceil log_2(|\mathcal{G}|) \rceil \cdot |\mathcal{G}|\cdot |\Sigma|^k}{8d} + \frac{|\mathcal{G}| \cdot \lceil log_2(|\Sigma|^k+1) \rceil}{8}
\label{e:dna_fm_size}
\end{equation}
where $k$ is the number of DNA symbols updated in each search iteration. The size of multi-step FM-Index exponentially increases with an enlarging $k$.

\begin{figure}[t!]
\centering
\includegraphics[width=3.2in]{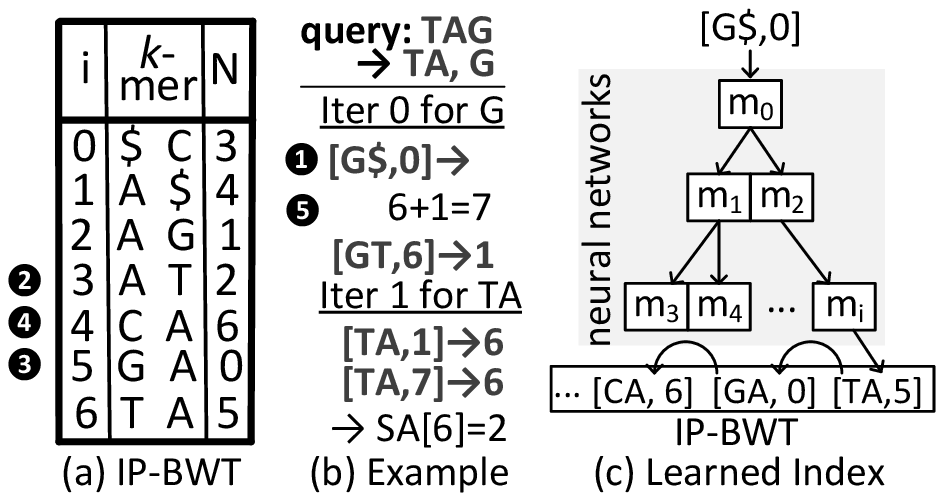}
\vspace{-0.15in}
\caption{LISA: (a) an IP-BWT array; (b) a search example; (c) a learned index.}
\label{f:dna_learned_index}
\vspace{-0.25in}
\end{figure}


\subsubsection{Learned Indexes for Sequence Analysis}

To support multi-step searches with smaller DRAM overhead, a recent work proposes Learned Indexes for Sequence Analysis (LISA)~\cite{Ho:WSMN2019} consisting of an Index-Paired BWT (IP-BWT) array and a learned index. 
\begin{itemize}[nosep,leftmargin=*]
\item \textbf{IP-BWT}. In Figure~\ref{f:dna_learned_index}(a), each entry of the IP-BWT is a pair of [$k$-mer, $N$], where $k$-mer is the first $k$ symbols of the corresponding BW-matrix row, and $N$ is the row number of the sequence with the first $k$ and the last $|\mathcal{G}|-k$ symbols swapped in the BW-matrix. For example, if $k=2$, the $k$-mer of the row 0 of the IP-BWT can be derived from the row 0 of the BW-matrix, $\$CATAGA$, shown in Figure~\ref{f:dna_fmindex_basic}(a) using only the first 2 symbols $\$C$. By sweeping the first 2 symbols and the other 5 symbols of $\$CATAGA$, we can have $ATAGA\$C$, which is the row 3 of the BW-matrix. So the row 0 of the IP-BWT is [$\$C$, 3].

\item \textbf{Backward Search}. The backward search of LISA finds the lower bound position of a [$k$-mer, $N$] pair in the IP-BWT. Since the IP-BWT is sorted, LISA adopts binary search for backward searches. As Figure~\ref{f:dna_learned_index}(b) shows, to search the query $TAG$, we first break it into $TA$ and $G$, since each iteration can process a 2-mer. In the first iteration, we start with $G$. The padding algorithm~\cite{Ho:WSMN2019} of LISA converts $G$ to $G\$$ for $low$ and $GT$ for $high$. $low$ and $high$ are initialized to 0 and 6 respectively. \ding{182} To search [$G\$$, 0], a binary search is performed over the IP-BWT. \ding{183} During the binary search, $G\$$ is first compared against $AT$, i.e., the row 3 of the IP-BWT. \ding{184} Because $G\$>AT$, the binary search goes to the row 5, i.e., $GA$ of the IP-BWT. \ding{185} Finally, it ends with the row 4 of the IP-BWT, i.e., $CA$. \ding{186} Since $G\$>CA$, the new $low$ is calculated as $6+1=7$. $high$ can be computed in the same way. Each search iteration requires $log_2(|\mathcal{G}|)$ comparisons due to binary search.    

\item \textbf{Learned Index}. To reduce the number of comparisons during binary searches, LISA adopts a learned index~\cite{Kraska:ICMD2018}, i.e., a model hierarchy consisting of multiple neural network models, as shown in Figure~\ref{f:dna_learned_index}(c), where $m_i$ is the neural network model $i$. The learned index enables LISA to do only one comparison during each iteration in the best case. To search $[G\$, 0]$ by the learned index, we can traverse down lower-level neural network models based on the output of the higher-level neural network models. Finally, a leaf neural network model predicts the position of $[G\$, 0]$ in the IP-BWT. However, if the predicted position does not contain $[G\$, 0]$, a linear search over the IP-BWT starts from the predicted position to find its actual position. 
\end{itemize}

\begin{figure*}[t!]
\subfigure[Random FM-Index accesses]{
   \includegraphics[width=1.6in]{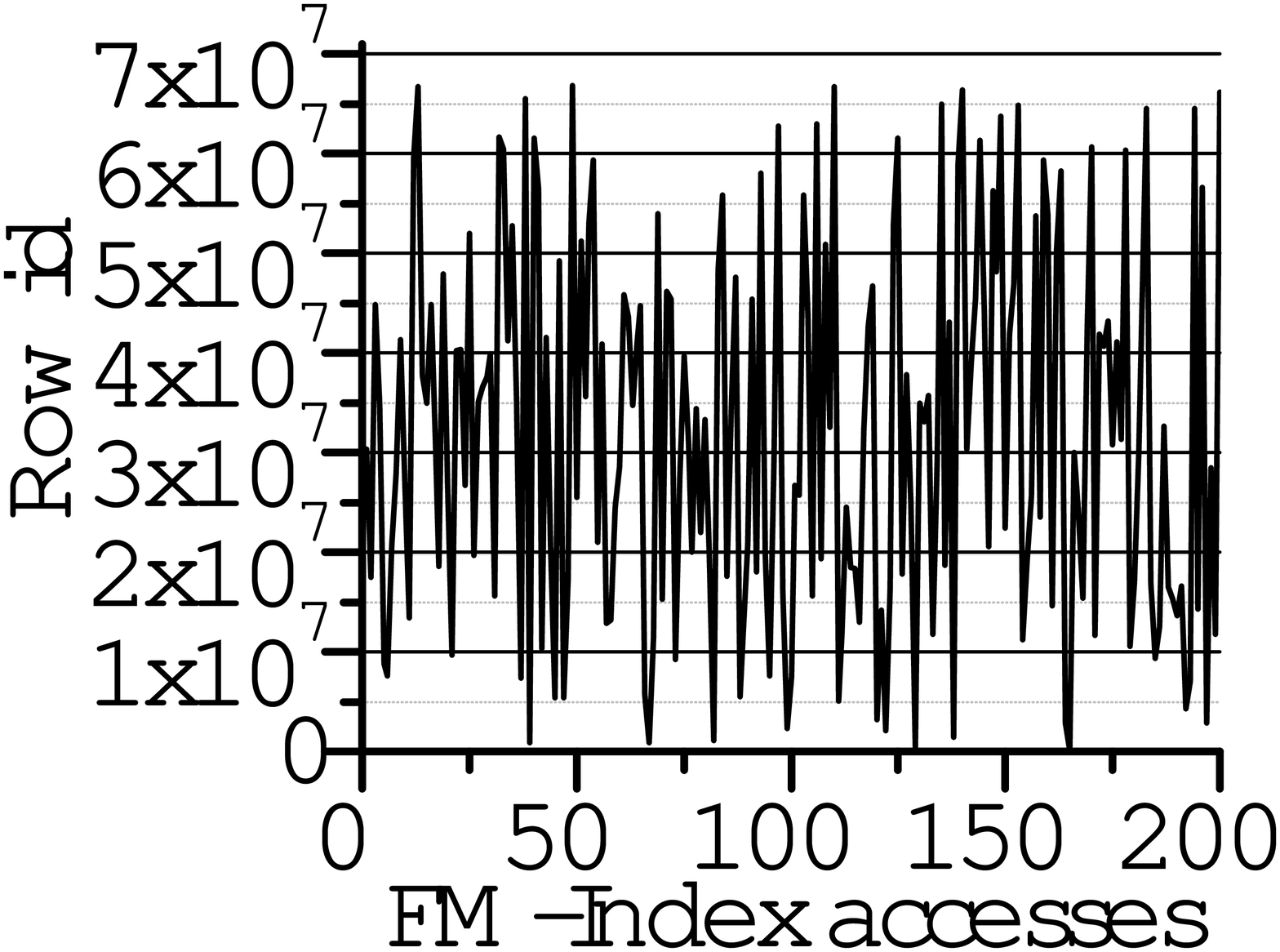}
   \label{f:dna_random_acc}
}
\subfigure[DRAM overhead v.s. step \#]{
   \includegraphics[width=1.6in]{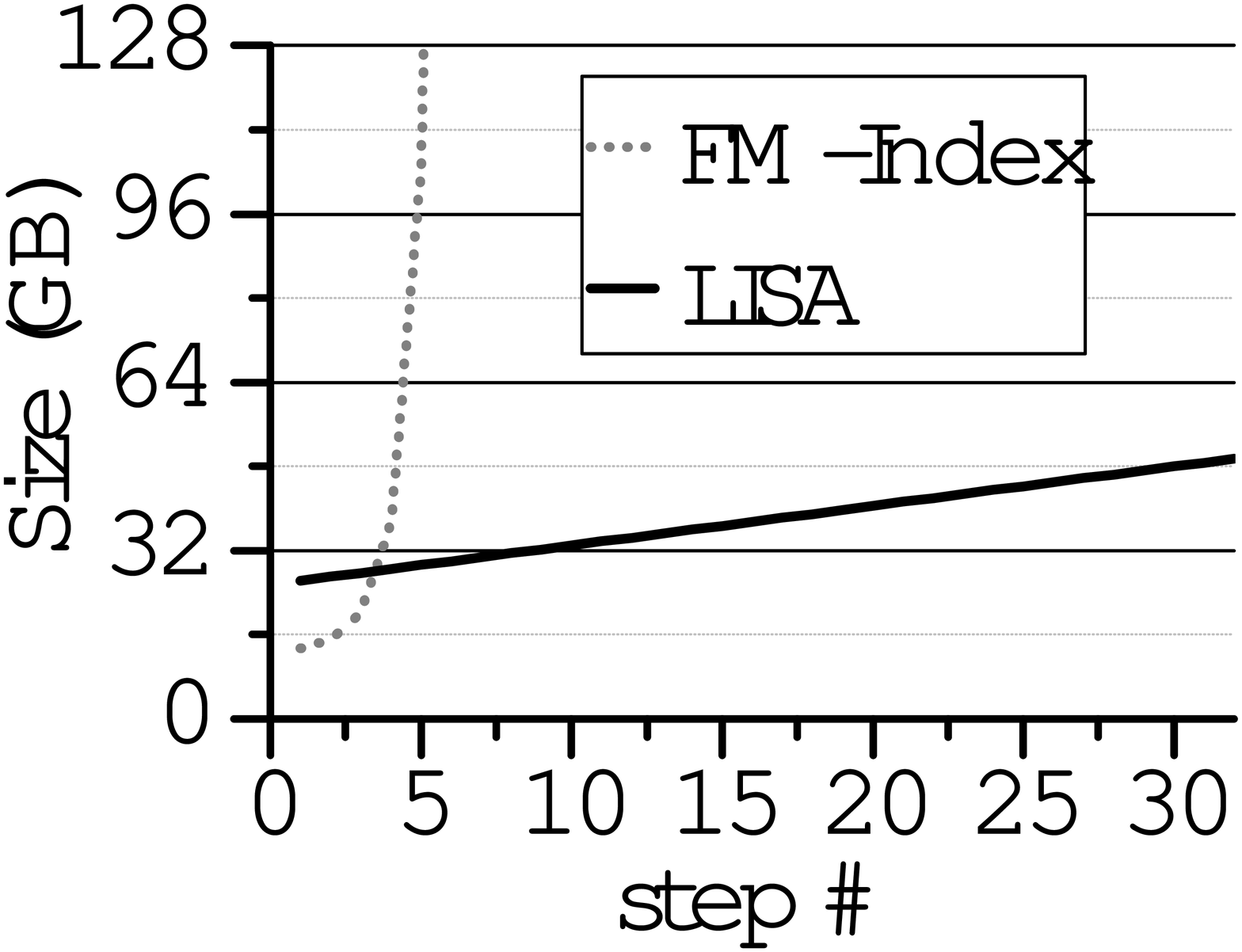}
   \label{f:dna_data_size}
}
\subfigure[Inaccurate LISA-21]{
   \includegraphics[width=1.6in]{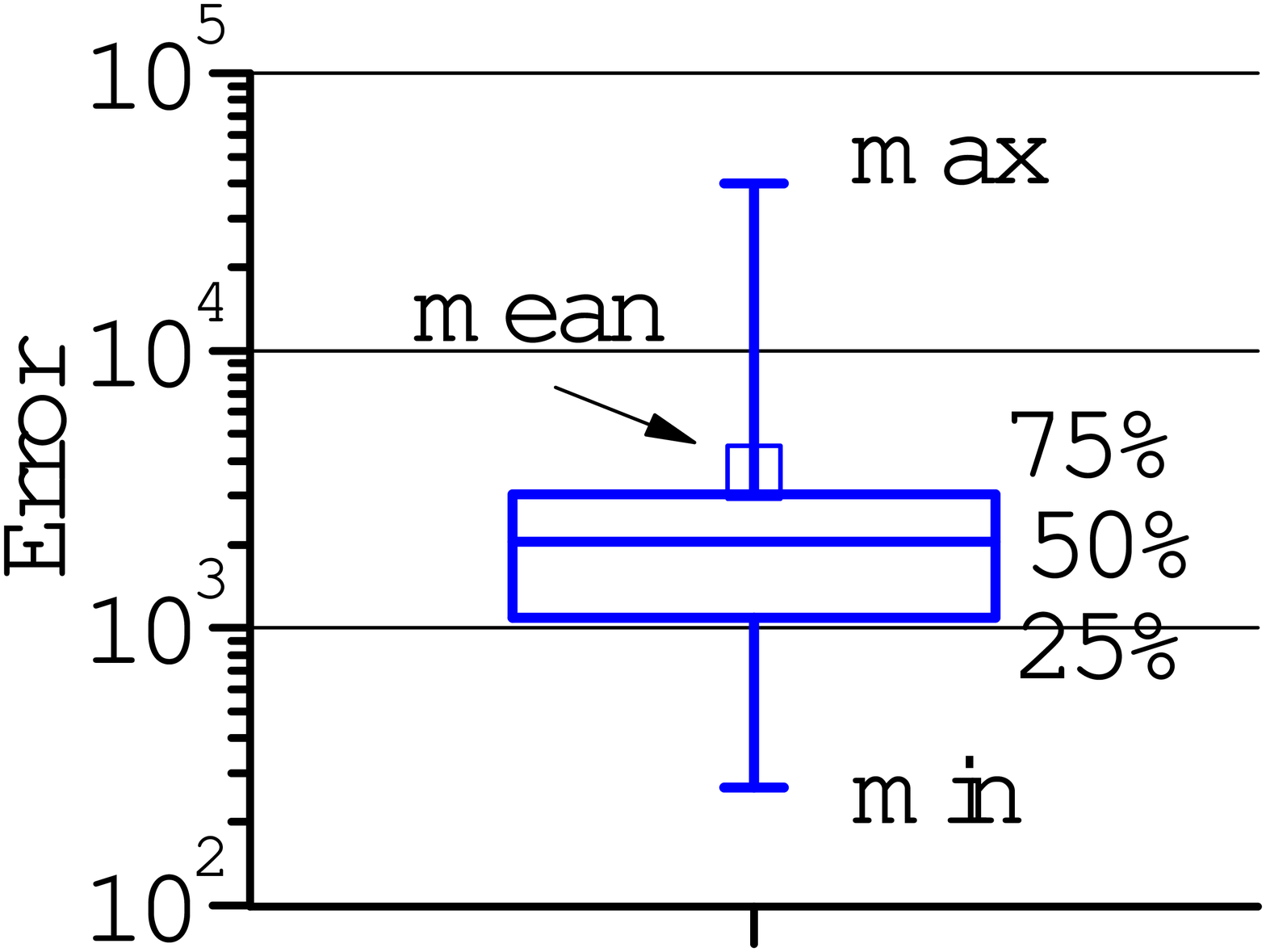}
   \label{f:dna_net_acc}
}
\subfigure[Throughput on CPU baseline]{
   \includegraphics[width=1.6in]{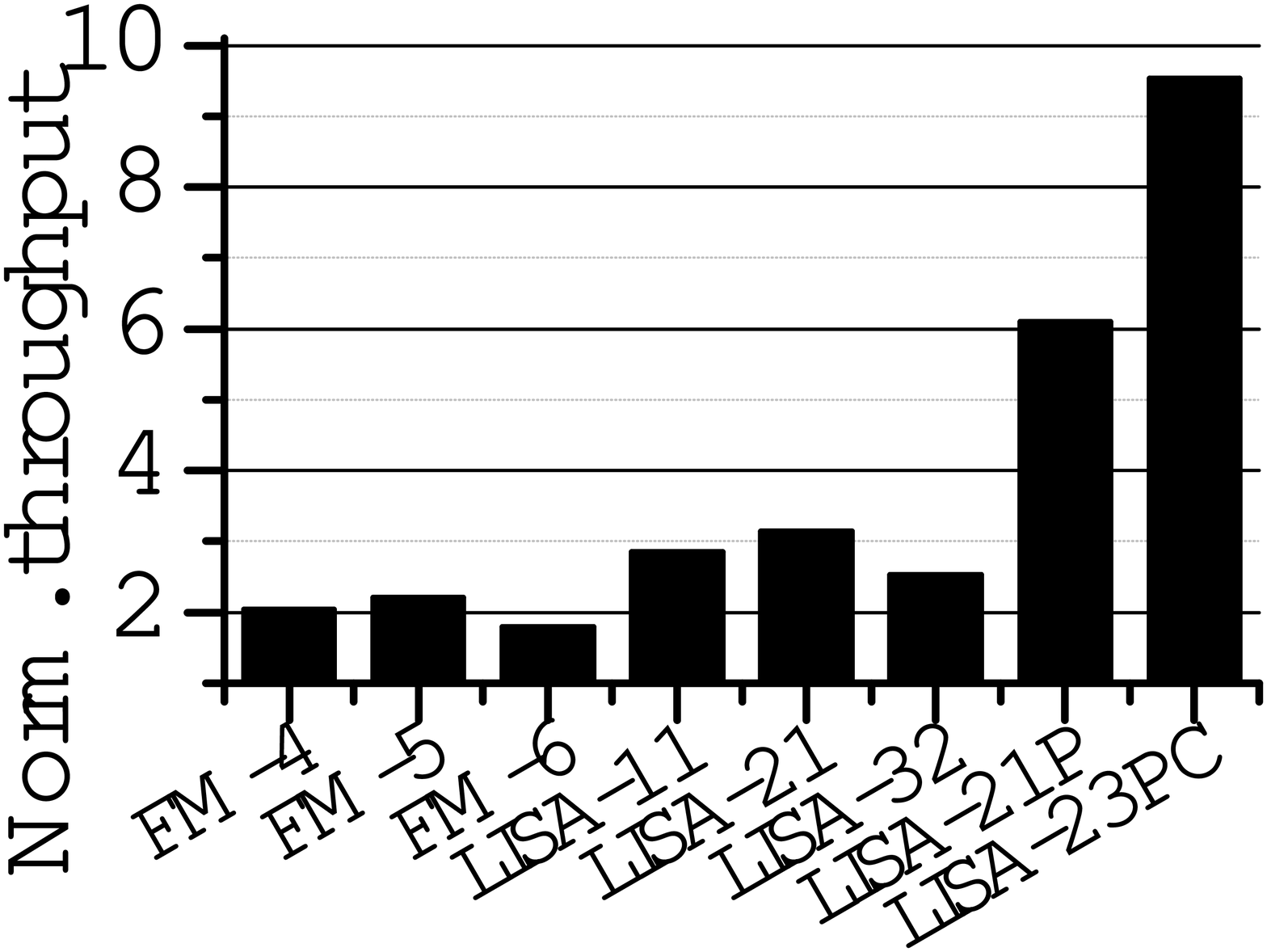}
   \label{f:dna_data_perf}
}
\vspace{-0.15in}
\caption{The inefficiency of prior FM-Index algorithms (A human genome dataset is used). In Figure~\ref{f:dna_net_acc}, we present the maximal and minimal errors, the mean of errors, and the $25_{th}$, $50_{th}$, $75_{th}$ percentiles of errors. In Figure~\ref{f:dna_data_perf}, \texttt{FM-X} means X-step conventional FM-Index; \texttt{LISA-X} means X-step LISA; \texttt{LISA-XP} means X-step LISA with a 100\% accurate index; and \texttt{LISA-XPC} denotes X-step LISA with a 100\% accurate index and a 100\% hit cache.}
\label{f:dna_weak_algorithm}
\vspace{-0.2in}
\end{figure*}

\section{Related Work and Design Motivation}
\label{s:moti}

It is challenging to achieve high-throughput and power-efficient FM-Index searches by state-of-the-art FM-Index algorithms and accelerators. 

\subsection{Algorithm Inefficiency}

\textbf{Intractable Size of $k$-step FM-Index}. We recorded the row IDs of 200 consecutive 1-step FM-Index search iterations in Figure~\ref{f:dna_random_acc}, where 197 different rows are accessed. Because 1-step FM-Index processes only 1 DNA symbol in each iteration, in most cases, searching a DNA symbol by 1-step FM-Index requires one row activation. Prior accelerators~\cite{Huangfu:MICRO2019,Zokaee:PACT2019} for 1-step FM-Index expect no row buffer hit and thus adopt close-page policy. Though $k$-step FM-Index can search $k$ DNA symbols by activating a row, as Figure~\ref{f:dna_data_size} shows, its data structure size exponentially increases with the step number $k$. For instance, 5-step FM-Index costs 105GB, while 6-step FM-Index occupies 374GB. As a result, 5-step FM-Index (\texttt{FM-5}) improves search throughput by only $1.21\times$ over 1-step FM-Index, as shown in Figure~\ref{f:dna_data_perf}. Further enlarging the step number of $k$-step FM-Index (\texttt{FM-6}) decreases its search throughput improvement, due to more TLB misses.

\textbf{Weakness of LISA Learned Index}. LISA can search $k$ DNA symbols after each row activation by its IP-BWT. Moreover, as Figure~\ref{f:dna_data_size} shows, the size of LISA linearly increases with the step number $k$. However, the learned index of LISA suffers from \textit{low accuracy} and \textit{low cache hit rate}. First, the LISA learned index has to index $|\mathcal{G}|$ IP-BWT entries, where $|\mathcal{G}|$ is the length of reference genome. For a human genome, $|\mathcal{G}|=3G$. When the learned index of LISA predicts a wrong position, LISA has to linearly search up to $|\mathcal{G}|$ IP-BWT entries. As Figure~\ref{f:dna_net_acc} describes, on average, LISA has to search $\sim3K$ extra IP-BWT entries during each iteration, due to the low accuracy of its learned index. Consequently, compared to 1-step FM-Index, 21-step LISA (\texttt{LISA-21}) improves search throughput by only $2.15\times$ in Figure~\ref{f:dna_data_perf}. But if LISA has a perfect learned index (\texttt{LISA-21P}) which always predicts the right position, compared to 1-step FM-Index, \texttt{LISA-21P} improves search throughput by $5.1\times$. Further increasing the step number of LISA (\texttt{LISA-32}) also introduces more TLB misses, thereby achieving smaller search throughput improvement. Second, traversing the learned index's hierarchical models is also pointer chasing. The LISA learned index consumes $\sim1.5GB$. If we assume a perfect cache (100\% hit) for \texttt{LISA-21P} (\texttt{LISA-21PC}), \texttt{LISA-21PC} improves search throughput by $8.53\times$ over 1-step FM-Index.

\textbf{Algorithmic Takeaways}. \textit{(1) Implementing $k$-step search with moderately enlarged data structure is important for FM-Index. However, a larger step number, i.e., $k$, may not necessarily lead to better search throughput. (2) A more accurate learned index reduces linear search overhead to improve search throughput. (3) A higher cache hit rate of learned index also improves search throughput by reducing redundant memory accesses to learned index.}


\begin{figure}[t!]
\centering
\includegraphics[width=3.3in]{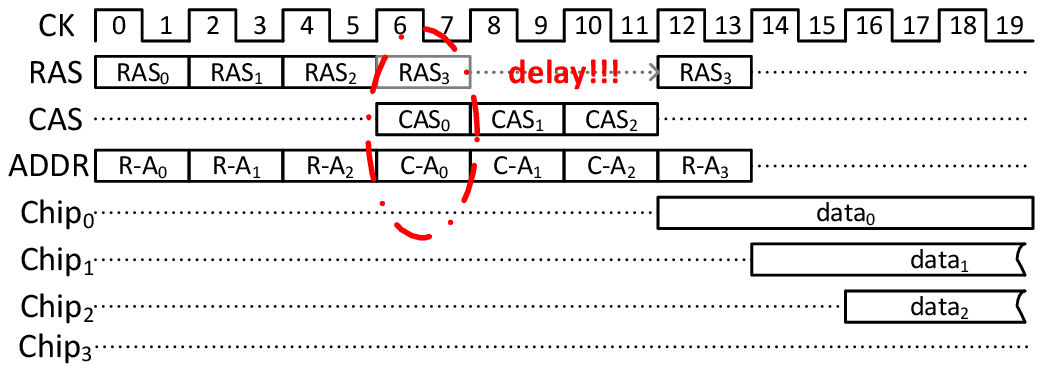}
\vspace{-0.15in}
\caption{The address bus bottleneck of MEDAL.}
\label{f:dna_medal_weak}
\vspace{-0.25in}
\end{figure}

\subsection{Hardware Incapacity}

\textbf{FPGAs and ASICs}. Most prior FM-Index hardware artifacts such as CPUs~\cite{Li:BWAMEM2013}, GPUs~\cite{Luo:PLOS2013}, FPGAs~\cite{Chang:FCCM2016}, and ASICs~\cite{Wang:ICPP2018} can accelerate only 1-step FM-Index searches. A recent FPGA design~\cite{Arram:TCBB2017} supports 2-step FM-Index searches, while $k$-step LISA is built merely on CPUs. Existing FM-Index application-specific accelerators including FPGAs~\cite{Chang:FCCM2016,Arram:TCBB2017}, and ASICs~\cite{Wang:ICPP2018} can search only 1 or 2 DNA symbols after each DRAM row activation. Therefore, they cannot fully exploit the maximal DRAM bandwidth.

\textbf{Processing-In-Memories}. Though recent works~\cite{Huangfu:MICRO2019,Zokaee:PACT2019,Shaahin:DAC2019,Angizi:DATE2020} propose PIM accelerators to process FM-Index searches, they cannot fully utilize the available DRAM bandwidth either. Most NVM-based PIMs~\cite{Zokaee:PACT2019,Shaahin:DAC2019,Angizi:DATE2020} focus on processing simple arithmetic computations of FM-Index by NVM arrays, but ignore optimizing external memory accesses. For instance, a ReRAM-based FM-Index PIM, FindeR~\cite{Zokaee:PACT2019}, has only 2.6GB memory arrays, so it still suffers from low DRAM bandwidth utilization when fetching FM-Index buckets that cannot fit into its internal arrays from external DRAMs. Only a DRAM PIM, MEDAL~\cite{Huangfu:MICRO2019}, modifies its DRAM main memory for higher FM-Index search throughput. Instead of processing multiple DNA symbols during a DRAM row activation, MEDAL enables chip-level parallelism where each chip in a rank can independently process a DNA symbol by opening its partial row. In ideal case, 16 chips in a rank can enlarge FM-Index search throughput by $16\times$, and each chip has only 1/16 row size. However, we find the DDR4 address bus shared by all chips in a rank seriously limits search throughput of MEDAL. The DDR4 address bus is only 17-bit wide~\cite{JEDEC:JESD79-4C}. During each access, the row activation and the column access serially pass their addresses via the same 17-bit bus. As Figure~\ref{f:dna_medal_weak} shows, MEDAL can sequentially activate a 1/16 partial row in $chip_{0\sim2}$. But when activating a 1/16 partial row in $chip_3$, its row address ($R$-$A_3$) and the column address of $chip_0$ ($C$-$A_0$) compete for the same address bus. The activation in $chip_3$ has to be delayed to $CK_{12}$. And idle bubbles appear in the DDR4 data bus. Because of address bus conflicts, although MEDAL claims a $68\times$ search throughput improvement over a multi-core CPU baseline, we observe it improves search throughput by only $11\times$.

\begin{figure}[t!]
\centering
\includegraphics[width=2.5in]{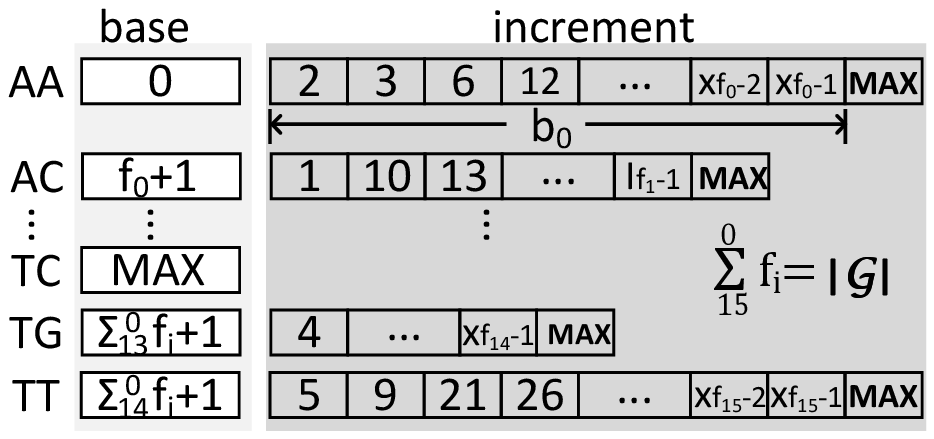}
\vspace{-0.1in}
\caption{A 2-step EXMA table ($MAX$ means the end of increments of a $k$-mer; and $f_i$ indicates the number of increments of the $i_{th}$ $k$-mer).}
\label{f:dna_exma_index}
\vspace{-0.2in}
\end{figure}

\section{EXMA}

We first create a row-buffer-friendly alternative to FM-Index, EXMA table, to process multiple DNA symbols in each search iteration. And then, we present a Multi-task-Learning (MTL)-based index to accelerate searches over an EXMA table. Compared to LISA learned index, the MTL-based index is more accurate, although it has less parameters.


\subsection{EXMA Table}

\textbf{Data Structure}. The major reason why the learned index of LISA is not accurate is that it has to index $|\mathcal{G}|$ IP-BWT entries. To reduce the problem size for a learned index, we propose a novel data structure, EXMA table. In each row of the $Occ$ table, only one $k$-mer can increase its value by one. For instance, in the 2-step $Occ$ table shown in Figure~\ref{f:dna_occ_mult88}, only ``AC'' increases its value from 0 to 1 in the row 1. Based on this observation, as Figure~\ref{f:dna_exma_index} shows, for each $k$-mer, an EXMA table stores only the row numbers of the $Occ$ table, where its value increases. For example, for ``AA'', its value increases in the row 2, 3, 6, and others. We store these row number as the \textit{increment}s of ``AA''. Totally, we have $f_0$ ``AA''s, $f_1$ ``AC''s, $\ldots$, and $f_{15}$ ``TT''s in the $Occ$ table, where $f_0,\dots,f_{15}$ are non-negative integers and $\sum_{i=0}^{15}f_i=|\mathcal{G}|$. So in the EXMA table, each 2-mer has $f_i$ increments, where $0\leq i \leq 15$. We add a symbol $MAX$ to indicate the end of the increments of a $k$-mer, where $MAX=|\mathcal{G}|+1$. The increments of an EXMA table has the space complexity of $\mathcal{O}(|\mathcal{G}|log(|\mathcal{G}|))$. For a 3G-base human genome, the increments occupy 12GB. We can consecutively store the increments of all $k$-mers to take advantage of the row buffer locality. Each $k$-mer needs a \textit{base} to point to its first increment. For instance, the base of ``AC'' is $f_0+1$ indicating its first increment is in the position $f_0+1$. Totally, a $k$-step EXMA table stores $4^k$ bases, each of which is related to a $k$-mer. Even if a $k$-mer has no increment, e.g., ``TC'', its base is set to $MAX=|\mathcal{G}|+1$. The bases of a $k$-step EXMA table have the space complexity of $\mathcal{O}(4^klog(|\mathcal{G}|))$.

\textbf{Backward Search}. Each iteration in a backward search computes Equation~\ref{e:dna_fm_lpm}. The entire $Count$ table costs only several bytes, so the bottleneck is the $Occ$ table lookups. To compute $Occ(k\mbox{-}mer, pos)$ in each search iteration, we compare $pos$ against all increments of the $k$-mer and find the first increment larger than $pos$, where $pos$ can be $low$ or $high$. For instance, to compute $Occ(AA, 4)$, we first read the base of ``AA'', which is 0. And then, we initialize a counter and start a linearly search from the position 0 of ``AA'' to $MAX$. If an increment is smaller than 4, we increase the counter by 1. When 6 is found, we stop, since $6>4$. At last, the counter value is 2, i.e., $Occ(AA, 4)=2$.

\begin{figure}[t!]
\centering
\subfigure[Learned index]{
   \includegraphics[width=1.8in]{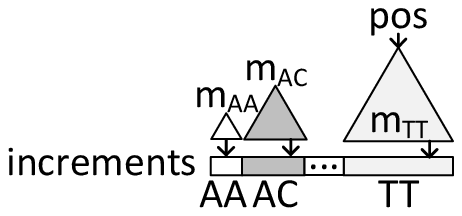}
   \label{f:dna_hash_tt}
}
\subfigure[MTL index]{
   \includegraphics[width=1in]{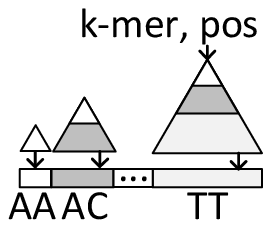}
   \label{f:dna_mtl_index2}
}
\vspace{-0.1in}
\caption{MTL-based EXMA index.}
\label{f:dna_hash_table2}
\vspace{-0.3in}
\end{figure}

\textbf{Na\"ive Adoption of Learned Index}. Since all increments of each $k$-mer are sorted, similar to LISA, we can adopt learned index~\cite{Kraska:ICMD2018} to perform only one comparison to compute $Occ(k\mbox{-}mer, pos)$ in the best case. We build a learned index model hierarchy for each $k$-mer that has $>256$ increments. As Figure~\ref{f:dna_hash_tt} shows, similar to LISA~\cite{Ho:WSMN2019}, to build learned indexes, we use a fixed ratio between the number of parameters of a learned index model and the number of increments that need to be indexed. As a result, if a $k$-mer has more increments, its learned index model has more parameters. For instance, the model of  ``TT'' ($m_{TT}$) owns more weights and biases than that of ``AA'' ($m_{AA}$), since ``TT'' has more increments. To compute $Occ(AA, pos)$ in a search iteration, we first read all the parameters of $m_{AA}$, and input only $pos$ to the model. If the prediction is not correct, we start a linear search from the predicted position to find the correct position. However, since an EXMA table have to index totally $|\mathcal{G}|$ increments, the learned index of EXMA does not have higher accuracy than that of LISA.

\begin{figure}[t!]
\centering
\subfigure[DRAM overhead v.s. step \#]{
   \includegraphics[width=1.55in]{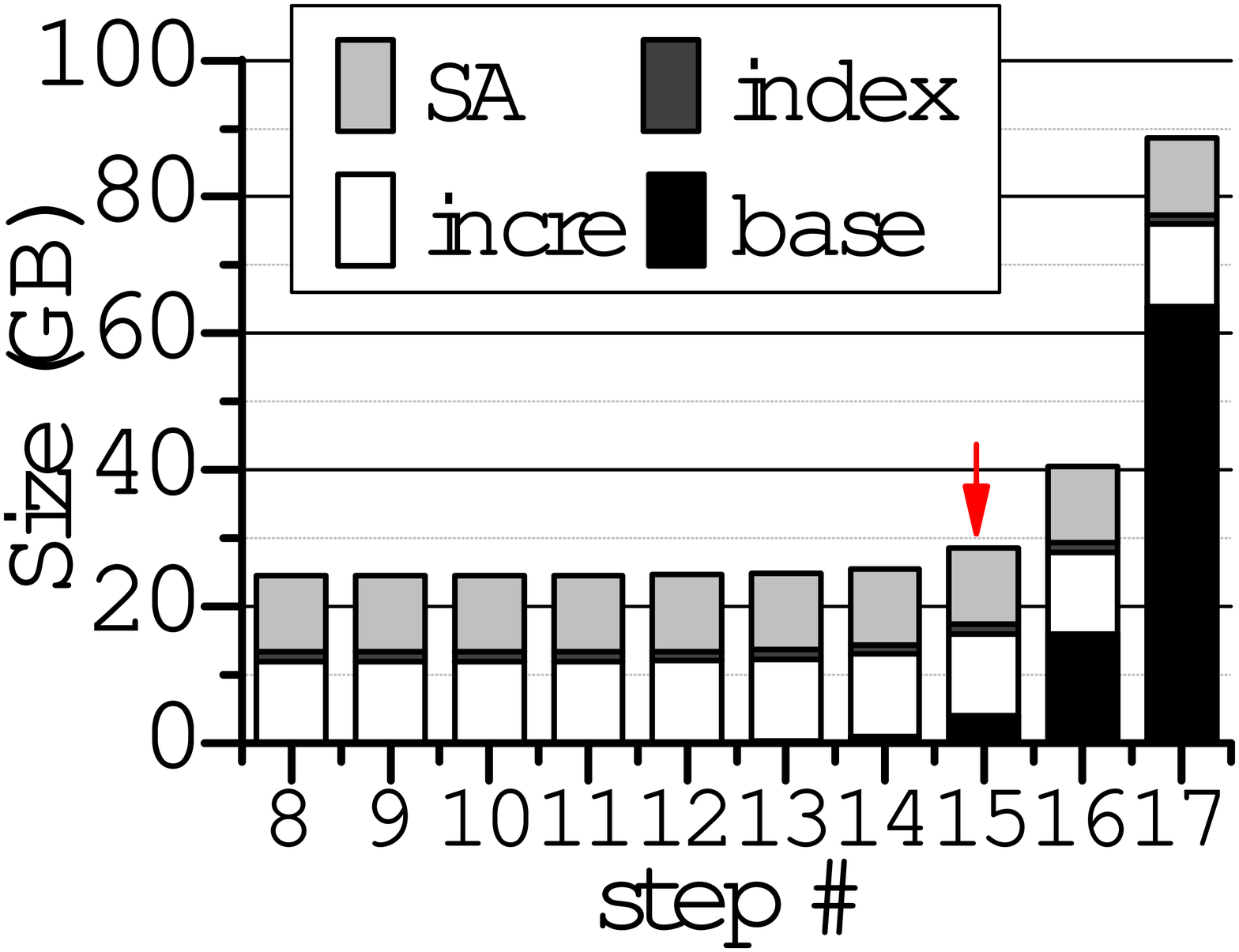}
   \label{f:dna_exma_data}
}
\hspace{-0.1in}
\subfigure[Throughput on CPU baseline]{
   \includegraphics[width=1.55in]{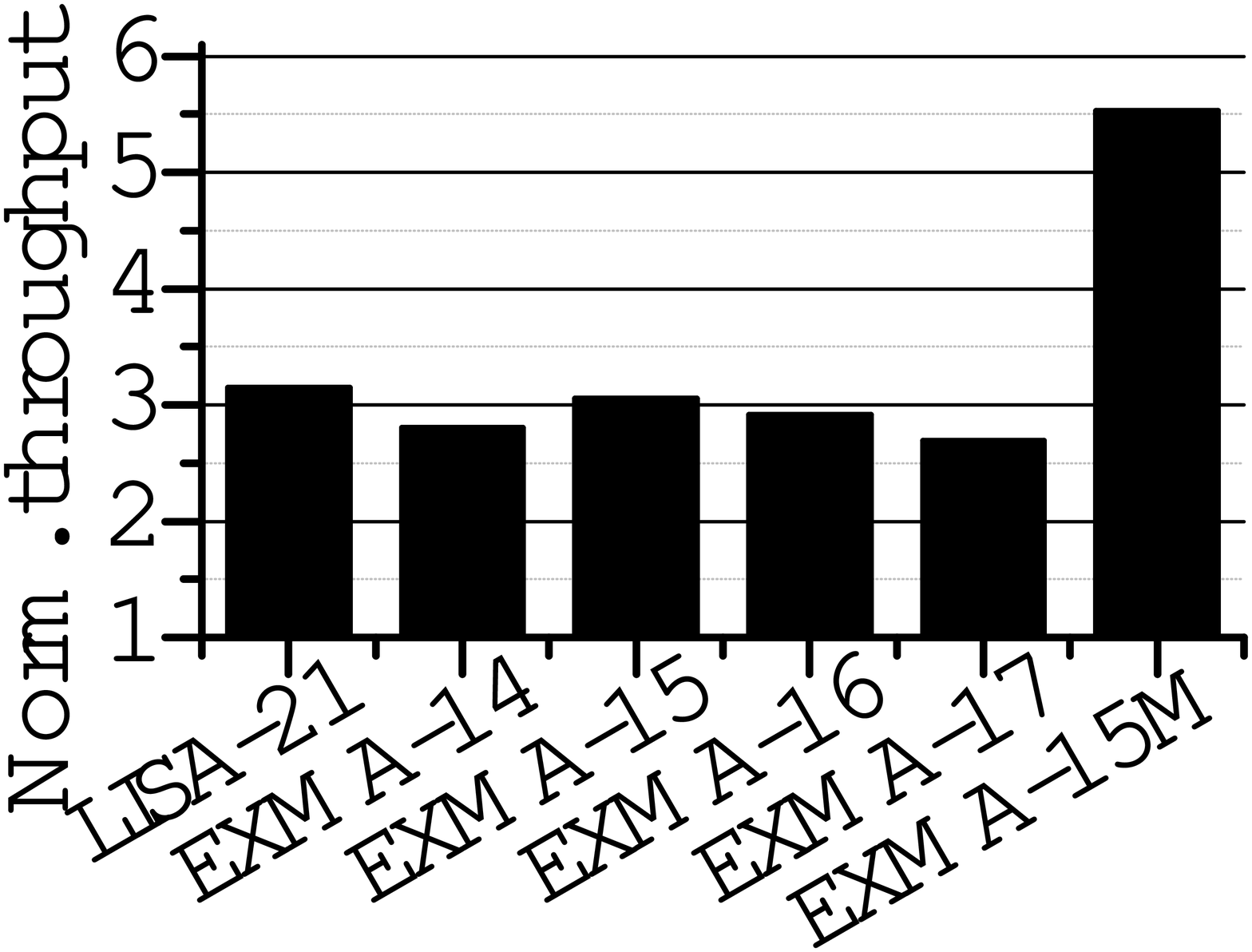}
   \label{f:dna_exma_perf}
}
\vspace{-0.1in}
\caption{The trade-off of an EXMA table (EXMA-X means X-step EXMA).}
\label{f:dna_exma_all}
\vspace{-0.1in}
\end{figure}

\textbf{Step \# of an EXMA Table}. We tuned the step number $k$ of an EXMA table to balance the DRAM overhead and search throughput. For a genome reference $\mathcal{G}$, the size of increments in an EXMA table is constant, while the size of bases in the table is proportional to $4^k$. Although a small $k$ has few bases, the search throughput is low. For instance, for a 3G-base human genome, if $k=2$, the bases require only 32-byte. But each time, only a 2-mer can be processed by a search iteration. In contrast, a large $k$ improves the search throughput but significantly increases the number of bases and thus the size of an EXMA table. For a human genome, as Figure~\ref{f:dna_exma_data} shows, a 15-step EXMA table (\texttt{15}) costs 29.5GB, while a 16-step EXMA (\texttt{16}) occupies 41.5GB. Increasing $k$ from 15 to 16 increases 12GB DRAM overhead. As Figure~\ref{f:dna_exma_perf} describes, EXMA (\texttt{EXMA-15}) achieves its best search throughput with 15-step. Compared to \texttt{LISA-21}, \texttt{EXMA-15} degrades search throughput by 7.3\%, since it has a smaller step number.

\begin{figure}[t!]
\centering
\subfigure[15 ``A''s]{
   \includegraphics[width=0.95in]{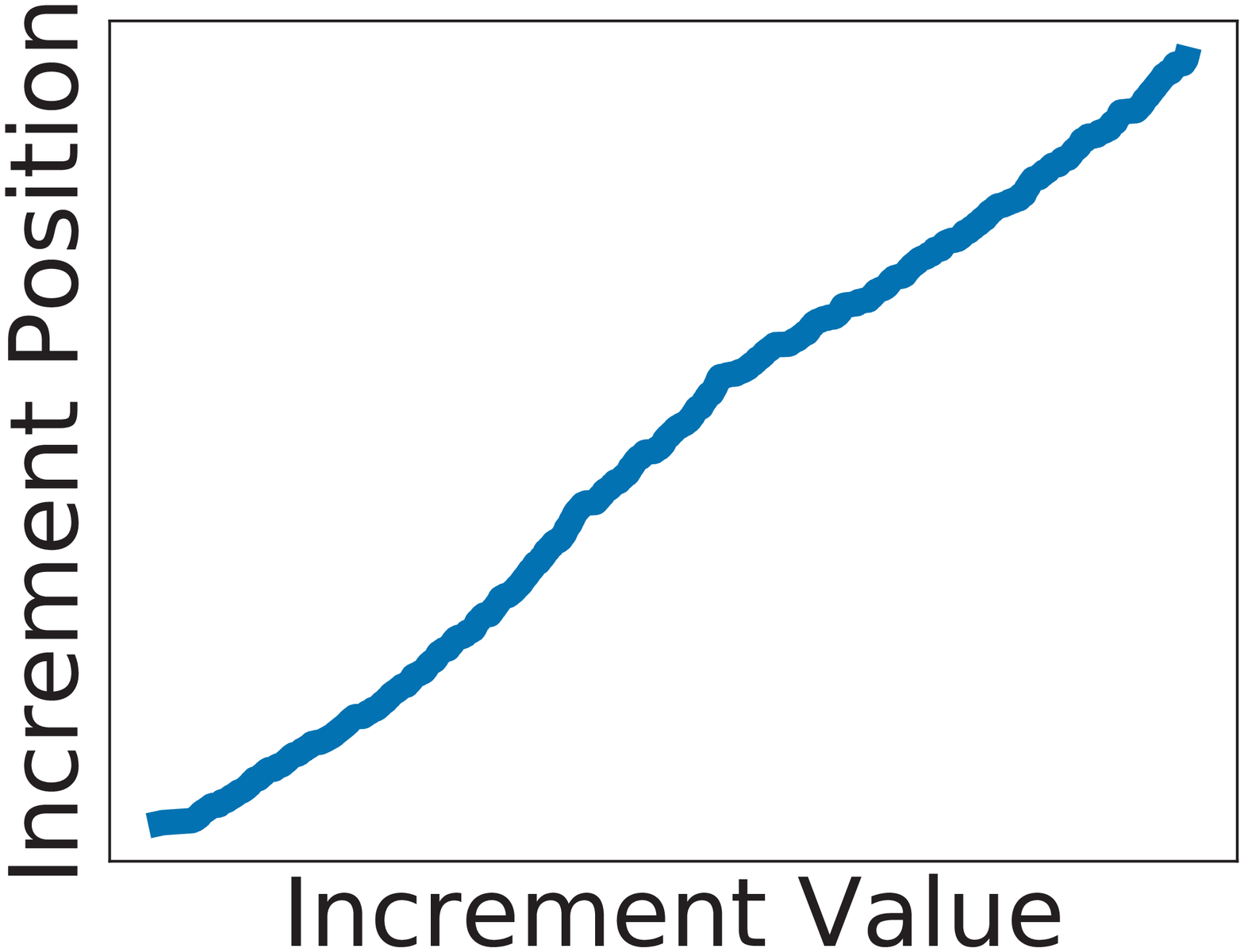}
   \label{f:incre_1}
}
\hspace{-0.1in}
\subfigure[7 ``AC''s and ``A'']{
   \includegraphics[width=0.95in]{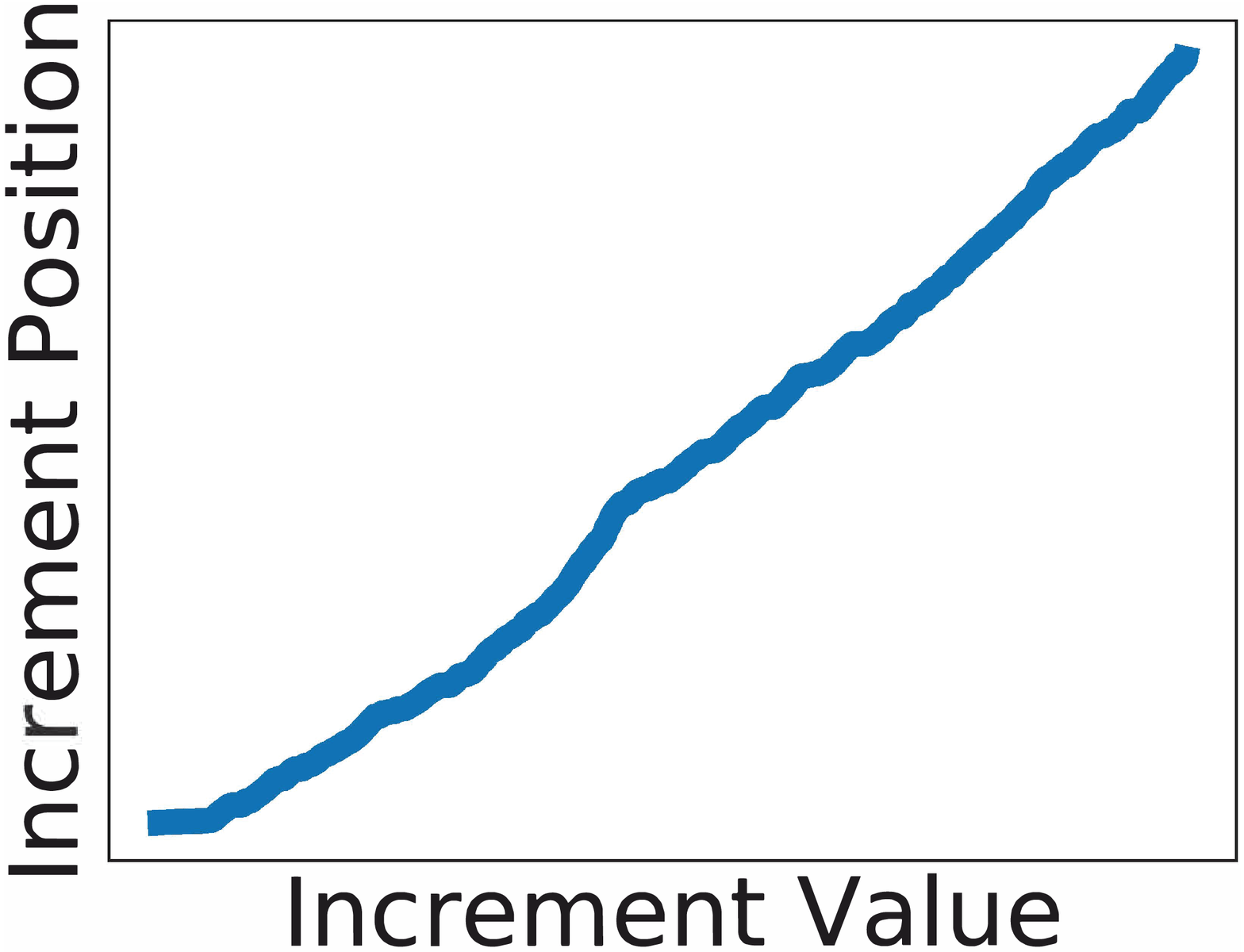}
   \label{f:incre_2}
}
\hspace{-0.1in}
\subfigure[7 ``AT''s and ``G'']{
   \includegraphics[width=0.95in]{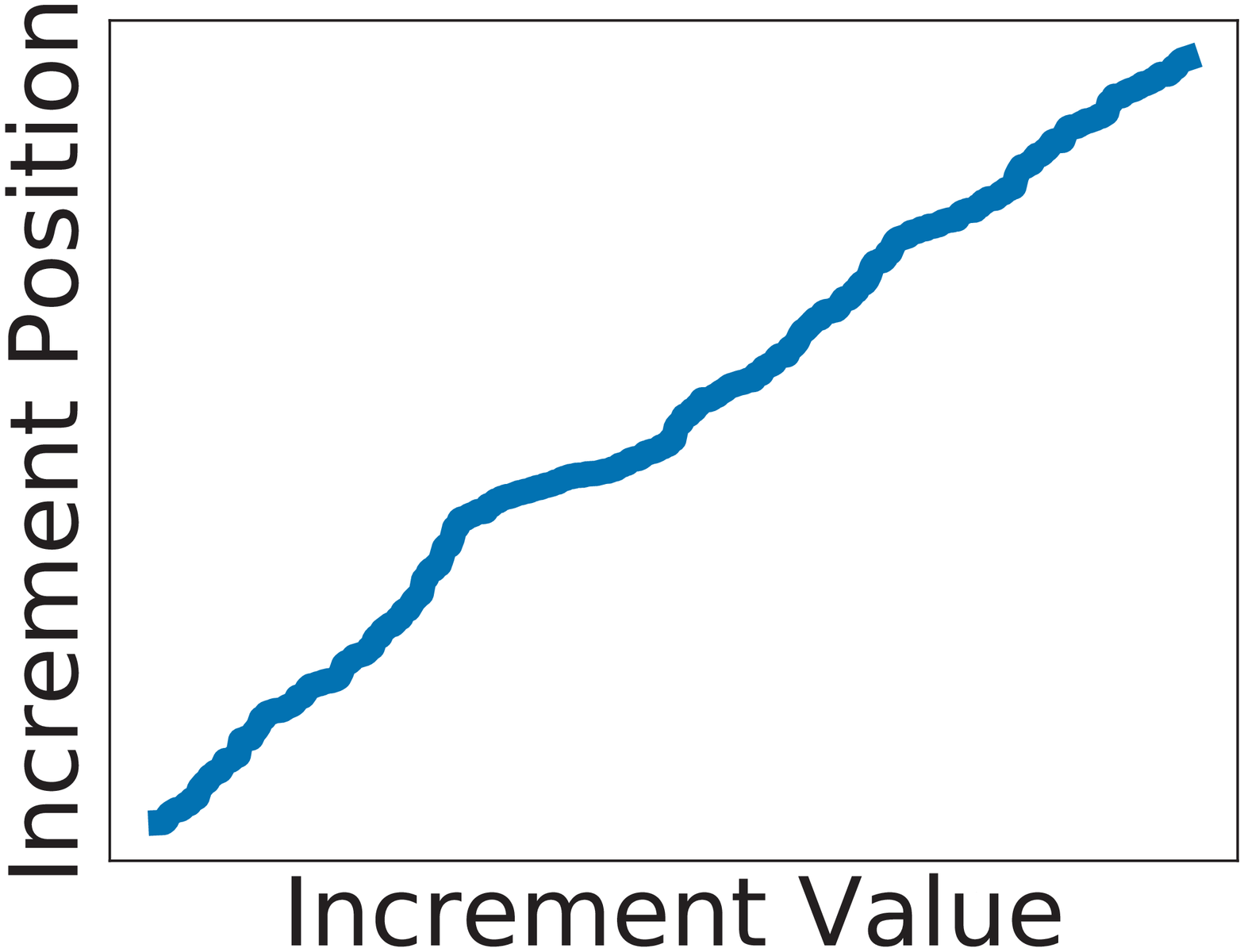}
   \label{f:incre_3}
}
\vspace{-0.1in}
\caption{Increment distributions of 15-mers.}
\label{f:dna_all_baba}
\vspace{-0.25in}
\end{figure}

\subsection{Multi-task-Learning Index for EXMA}
\label{s:mtlindex}

\textbf{MTL-based Index}. To more accurately approximate the cumulative distribution function of increments of each $k$-mer in an EXMA table with less parameters, we propose a Multi-Task-Learning~\cite{Ruder:ARXIV2017,Kendall:CVPR2018,Bilen:NIPS2016} (MTL)-based index for the increments of an EXMA table. As Figure~\ref{f:dna_all_baba} shows, the increments of various $k$-mers in \texttt{15-EXMA} exhibit similar random distributions. Based on Stein's paradox~\cite{Stein:Stan1956}, it is more accurate to approximate many independent random distributions using samples from all of them rather than approximating them separately. The MTL-based index is trained with the increments of multiple $k$-mers to obtain superior accuracy over learning the increments of each $k$-mer independently. We adopt the \textit{hard-parameter-sharing} MTL~\cite{Bilen:NIPS2016} that shares a subset of parameters between the learned index models of $k$-mers having different numbers of increments. For instance, as Figure~\ref{f:dna_mtl_index2} shows, ``AA'' uses the smallest model ($m_{AA}$) to index its increments, since it has the fewest increments among all 2-mers. ``TT'' has more increments, and thus a larger model ($m_{TT}$) which contains the entire $m_{AA}$. Besides $m_{AA}$, $m_{TT}$ comprises more levels of nodes to index its increments more accurately. Compared to the na\"ive learned index, we add more neurons in the hidden layers of each non-leaf node of a MTL-based index to accommodate two inputs, i.e., $pos$ and $k$-mer. But the size of a MTL-based index is smaller that of the na\"ive learned index, since the $k$-mers share most parameters of a MTL-based index.

\textbf{Inference}. The MTL-based index of an EXMA table effectively approximates $Occ(k\mbox{-}mer, pos)$ as
\vspace{-0.05in}
\begin{equation}
p=F(k\mbox{-}mer, pos)*f_{k\mbox{-}mer}
\vspace{-0.05in}
\end{equation}
where $p$ is the predicted position of $pos$ in the increments of the $k$-mer; $F(.,.)$ is the neural network model hierarchy to estimate the probability to observe an increment $\leq pos$; $f_{k\mbox{-}mer}$ is the number of increments of the $k$-mer, and can be stored along the $k$-mer base. After each inference, we read both $p$ and $p+1$ to check whether the prediction is correct. If not, we start a linear search to find the correct position. Therefore, \textit{the accuracy of a MTL-based index decides search throughput of FM-Index, but has no impact on the quality of final DNA mapping}. A MTL-based index model hierarchy is a tree structure. To keep the index size in check, we deploy simple linear regression models~\cite{Kutner:ALSM2005} as leaf nodes in the model hierarchy of the MTL-based index. A linear regression model contains only one weight and one bias. Each non-leaf node is a neural network having a fully-connected layer, each of which contains 10 neurons with sigmoid activation.


\textbf{Training}. The MTL-based index is built for the increments of $p$ $k$-mers $\{\mathcal{T}_i\}_{i=1}^p$. For a $k$-mer $\{\mathcal{T}_i\}$, its training dataset includes $f_i$ increments $\{\mathbf{x}_{i,j}\}_{j=1}^{f_i}$ and their positions $\{y_{i,j}\}_{j=1}^{f_i}$. The learning function for the $k$-mer $\mathcal{T}_i$ is defined as $\mathbf{w}_i^T\mathbf{x}+b_i$. Based on~\cite{Argyriou:NIPS2007,Kendall:CVPR2018,Bilen:NIPS2016}, we formulate the loss function to learn the relations between $k$-mers as 
\vspace{-0.05in}
\begin{equation}
\displaystyle \min_{\mathbf{W},\mathbf{b}} \sum_{i=1}^p\frac{\beta_i}{f_i}\sum_{j=1}^{f_i} l(\mathbf{w}_i^T\mathbf{x}_{i,j}+b_{i}, y_{i,j})
\label{e:all_in_one}
\vspace{-0.05in}
\end{equation}
where $\mathbf{W}=(\mathbf{w}_1,\ldots,\mathbf{w}_p)$; $\mathbf{b}=(b_1,\ldots,b_p)^T$; $l(\cdot,\cdot)$ means the cross-entropy loss function; $\beta_i$ is the importance of $k$-mer $\{\mathcal{T}_i\}$. We trained the MTL-based index to minimize this equation by an Adam optimizer. Similar to LISA~\cite{Ho:WSMN2019}, the training and testing of a MTL-based index use the same dataset, an EXMA table of a reference. Training a MTL-based index for a reference typically takes $1\sim 2$ days. Once a MTL-based index is trained, billions of exact match operations can happen on its reference.

\begin{figure}[t!]
\centering
\includegraphics[width=3.3in]{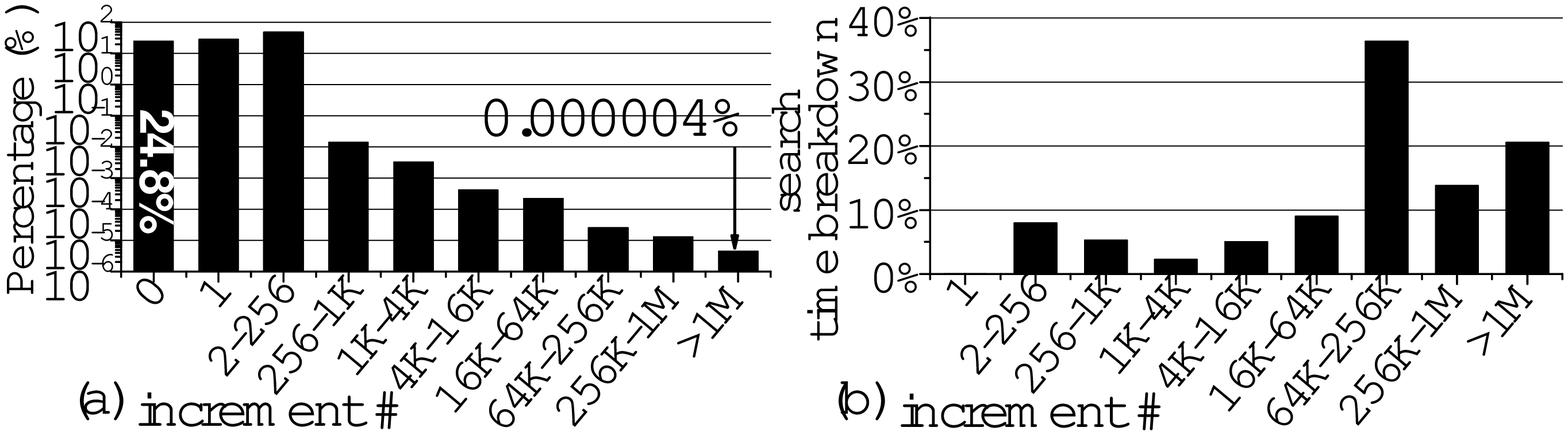}
\vspace{-0.1in}
\caption{Profiling \texttt{EXMA-15} : (a) increment \#; (b) search time breakdown.}
\label{f:dna_EXMA_all}
\vspace{-0.3in}
\end{figure}

\begin{figure}[h!]
\vspace{-0.2in}
\centering
\includegraphics[width=3.3in]{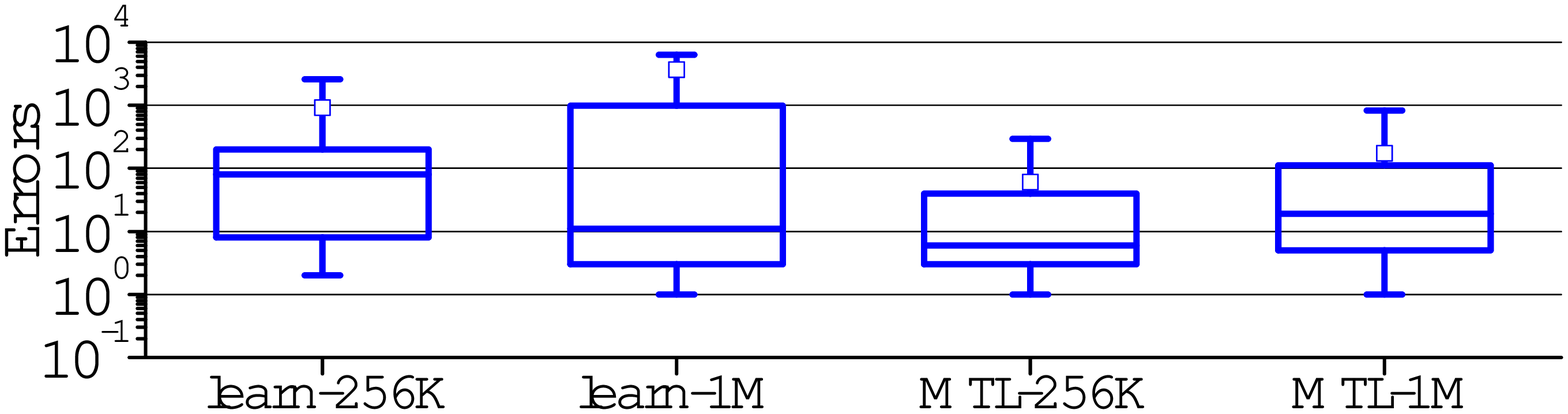}
\vspace{-0.1in}
\caption{Prediction errors of learned and MTL indexes.}
\label{f:dna_index_acc}
\vspace{-0.1in}
\end{figure}

\textbf{MTL-based Index Performance}. We profiled the perf-ormance of \texttt{EXMA-15} equipped with a na\"ive learned index in Figure~\ref{f:dna_EXMA_all}. As Figure~\ref{f:dna_EXMA_all}(a) shows, 2.5E-5$\%$ and 4E-6$\%$ of 15-mers have 64K$\sim$256K and $>$1M increments, respectively. However, searching the 15-mers having 64K$\sim$256K and $>$1M increments consumes 36\% and 20.5\% of the search time respectively, as shown in Figure~\ref{f:dna_EXMA_all}(b). This is because the na\"ive learned index predicts a lot wrong positions, and the linear search overhead is large. The prediction errors of the na\"ive learned index for the 15-mers having 64K$\sim$256K (\texttt{learn-256K}) and $>$1M (\texttt{learn-1M}) increments are shown in Figure~\ref{f:dna_index_acc}. On average, we have to search 917 and 2133 more increments to find the correct one for the 15-mers having 64K$\sim$256K and $>$1M increments respectively. The MTL-based index greatly improves index prediction accuracy by simultaneously learning from multiple 15-mers having similar amounts of increments. The MTL-based index (\texttt{MTL}) further reduces the mean of prediction errors to 45 and 182 for the 15-mers having 64K$\sim$256K and $>$1M increments respectively. As a result, the MTL-based index (\texttt{EXMA-15M}) improves search throughput by $75\%$ over \texttt{LISA-21} with only a half number of parameters, as shown in Figure~\ref{f:dna_exma_perf}.

\begin{figure}[t!]
\centering
\includegraphics[width=3.4in]{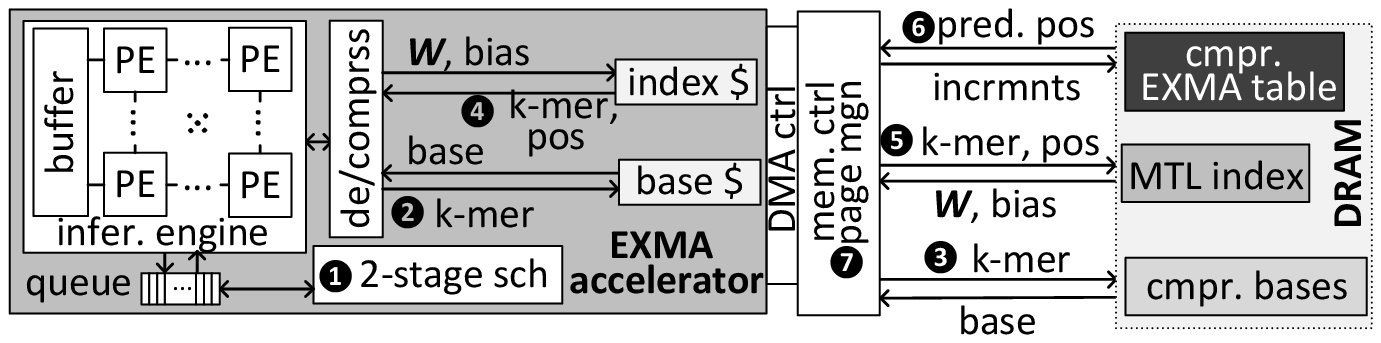}
\vspace{-0.15in}
\caption{The architecture of EXMA accelerator.}
\label{f:dna_fpga_arch}
\vspace{-0.3in}
\end{figure}

\subsection{EXMA Accelerator}

We propose a hardware accelerator to process search operations over an EXMA table using a MTL-based index. We integrate two caches for the bases and the MTL-based index of an EXMA table to reduce unnecessary DRAM accesses. And then, we present EXMA scheduling to improve cache hit rate. We also introduce dynamic page policy to improve DRAM row buffer hit rate during searching the increments of an EXMA table. At last, we create CHAIN compression to reduce the EXMA table size.

\begin{figure*}[t!]
\centering
\includegraphics[width=6in]{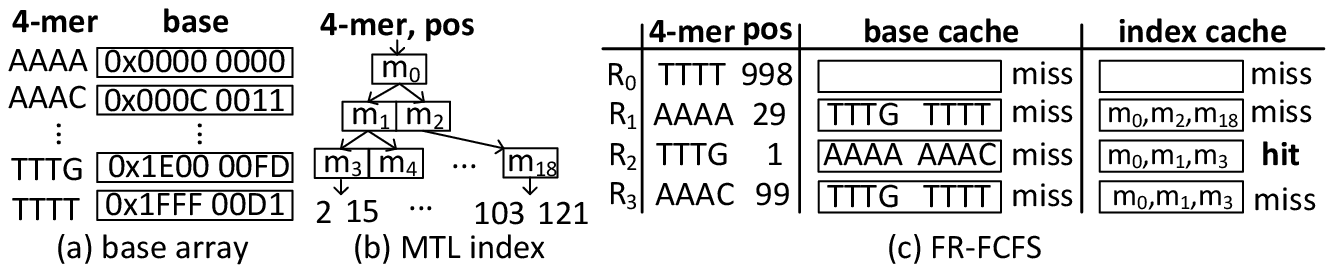}
\vspace{-0.15in}
\caption{The low cache hit rate of MTL-based index.}
\label{f:dna_low_locality}
\vspace{-0.2in}
\end{figure*}

\begin{figure}[t!]
\vspace{-0.1in}
\centering
\includegraphics[width=3in]{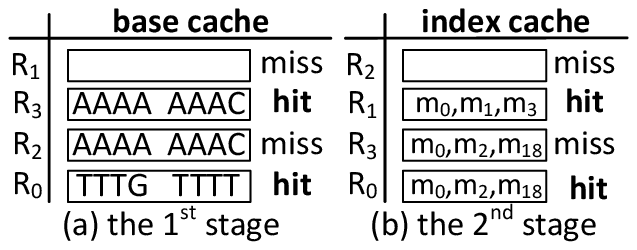}
\vspace{-0.15in}
\caption{The 2-stage EXMA scheduling.}
\label{f:dna_exma_scheduling}
\vspace{-0.3in}
\end{figure}

\subsubsection{Accelerator Architecture}

The architecture of our EXMA accelerator is shown in Figure~\ref{f:dna_fpga_arch}. The kernel of the EXMA accelerator is an inference engine computing predictions of a MTL-based index. We adopt the state-of-the-art Tangram neural network accelerator~\cite{Gao:ASPLOS2019} as the inference engine. The Tangram accelerator consists of a number of processing elements (PEs) organized in a 2D array. Each PE includes a simple ALU for multiply-accumulate (MAC) operations and a small register file of 32B. A larger SRAM buffer is shared by all PEs. We add two small caches for the bases and MTL-based index of an EXMA table. Data fetched and stored by the accelerator goes through a de/compression unit (\cref{s:chaincomp}) that de/compresses the bases and increments of an EXMA table. We integrate a scheduling queue into the EXMA accelerator to implement 2-stage EXMA scheduling (\cref{s:2-phasesche}). At last, the dynamic page management (\cref{s:dpp_all}) switching between open and close page policies is implemented in the CPU memory controller.

\subsubsection{EXMA Scheduling}
\label{s:2-phasesche}

\textbf{Poor Locality of MTL-based Index}. The conventional first-ready first-come-first-serve (FR-FCFS) policy is adopted by almost all accelerators~\cite{Huangfu:MICRO2019,Zokaee:PACT2019,Shaahin:DAC2019,Wang:ICPP2018,Chang:FCCM2016,Arram:TCBB2017} to schedule FM-Index searches. However, FR-FCFS significantly degrades the hit rates of our base cache and index cache. We show an example of FR-FCFS in Figure~\ref{f:dna_low_locality}, where there are 4 FM-Index requests in the form of $[k$-$mer, pos]$, i.e., $R_0=[TTTT, 998]$, $R_1=[AAAA, 29]$, $R_2=[TTTG, 1]$ and $R_3=[AAAC, 99]$. As Figure~\ref{f:dna_low_locality}(a) shows, all bases are stored consecutively in the lexicographical order of $k$-mers. Each base occupies 4-byte. So the bases of $AAAA$ and $AAAC$ ($TTTC$ and $TTTT$) are stored in the same 64-byte. The MTL-based index used by four requests is shown in Figure~\ref{f:dna_low_locality}(b). To predict the increments of $pos$ 1 and 29 (99 and 998), the MTL index nodes of $m_0$, $m_1$ and $m_3$ ($m_0$, $m_2$ and $m_{18}$) are required. Assume the base cache can contain only one 64-byte line, while the index cache can store three index nodes. Four requests are scheduled by FR-FCFS in Figure~\ref{f:dna_low_locality}(c). When $R_0$ arrives, both caches are empty and thus have a miss. And then, the bases of ``TTTG'' and ``TTTT'' are stored in the base cache, while the index nodes of $m_0$, $m_2$ and $m_{18}$ used by $R_0$ are stored in the index cache. For $R_1$, both caches also suffer from a miss. The bases of ``AAAA'' and ``AAAC'' are fetched to the base cache, while $m_0$, $m_1$ and $m_3$ used by $R_1$ are installed into the index cache. For $R_2$, the base cache has a miss, but the index cache has a hit. At last, both cache have a miss for $R_3$. Totally, four misses happen in the base cache, while three misses occur in the index cache.

\textbf{2-Stage Scheduling}. We propose a 2-stage EXMA scheduling technique to enhance the hit rates of the base and index caches. Unlike FR-FCFS scheduling requests based on their addresses and order, EXMA re-orders the requests according to their data including $k$-mers and positions ($pos$). In the first stage, EXMA sorts the FM-Index requests based on their $k$-mers. By consecutively issuing FM-Index requests in the lexicographic order of their $k$-mers, the hit rate of the base cache increases, since the bases of the $k$-mers sorted in the lexicographic order tend to have stronger spatial locality. As Figure~\ref{f:dna_exma_scheduling}(a) shows, the first stage of EXMA scheduling issues four requests in the order of $R_1$, $R_3$, $R_2$, and $R_0$. The base cache has two hits and two misses, but the index cache has all four misses. This is why EXMA needs to do the second stage of scheduling. During the second stage of EXMA scheduling, four requests are ranked according to their $pos$ values. Through consecutively computing inferences of MTL indexes of the requests having small differences between their $pos$ values, the index cache can expect more hits. This is because the smaller difference the $pos$ values of two requests exhibit, the more likely these two requests use the same MTL index nodes during searches. As Figure~\ref{f:dna_exma_scheduling}(b) shows, the index cache has two misses and two hits. Totally, the 2-stage EXMA scheduling has 4 misses and 4 hits.

\textbf{Implementation}. Our 2-stage EXMA scheduling is implemented with the scheduling queue that is a content-addressable memory (CAM). A CAM can perform sorting operations~\cite{Okabayashi:VLSI1990}. The $k$-mer and $pos$ of a request can be stored in one row of the CAM. Each DNA symbol in a $k$-mer is denoted by 3 bits, since we need to encode \$, A, C, T and G. Requests can be sorted in the CAM based on their $k$-mers or $pos$ values.

\subsubsection{Dynamic Page Policy}
\label{s:dpp_all}

\textbf{Dynamic Page Policy}. Prior FM-Index accelerators~\cite{Huangfu:MICRO2019,Zokaee:PACT2019,Shaahin:DAC2019,Wang:ICPP2018,Chang:FCCM2016,Arram:TCBB2017} adopt close-page policy in their DRAM main memories. They always pre-charge a DRAM row after each access, since conventional 1-step FM-Index searches have little spatial locality, as shown in Figure~\ref{f:dna_random_acc}. On the contrary, our EXMA table stores the increments of a $k$-mer consecutively in DRAM rows. As the algorithm of FM-Index backward search (line 2$\sim$3 in Figure~\ref{f:dna_fmindex_basic}(d)) indicates, each iteration issues two requests searching the same $k$-mer but with different position values. In a search iteration, we compute $Occ(k$-$mer,low)$ and $Occ(k$-$mer,high)$. Both search the increments of the same $k$-mer that are very likely stored in the same row. So our accelerator asks the CPU memory controller (MC) to keep the DRAM row open after the first request in a search iteration is processed, since we expect the second request can hit in the row buffer. However, the row will be pre-charged, after the second is processed.

\textbf{Implementation}. The dynamic page policy can be implemented in the CPU MC. When searching a request, if there is another request pending in the scheduling queue of the accelerator searches the same $k$-mer, the CPU MC keeps the DRAM row open after the ongoing request completes. Otherwise, it pre-charges that DRAM row. The CPU MC maintains a register to indicate whether all rows are closed, and another register to record which row is open for each bank.

\subsubsection{CHAIN compression}
\label{s:chaincomp}

\begin{figure}[t!]
\centering
\subfigure[B$\Delta$I]{
   \includegraphics[width=2.3in]{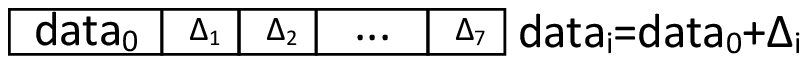}
   \label{f:dna_prior_com}
}
\subfigure[Compression and decompression of CHAIN]{
   \includegraphics[width=2.4in]{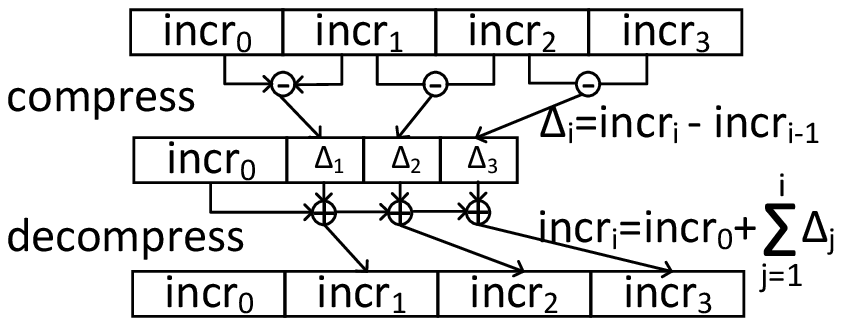}
   \label{f:dna_chain_all}
}
\vspace{-0.1in}
\caption{The CHAIN compression.}
\label{f:dna_chain_comp}
\vspace{-0.3in}
\end{figure}

\textbf{B$\Delta$I}. The state-of-the-art cache compression technique, B$\Delta$I~\cite{Pekhimenko:PACT2012}, breaks a 64-byte cache line into eight 8-byte data sections. As Figure~\ref{f:dna_prior_com} shows, to compress the cache line, B$\Delta$I records the first data section ($data_0$) and stores only the difference ($\Delta_i$) between each data section ($data_i$) and $data_0$. To decompress a data section, B$\Delta$I simply calculates $data_i=data_0+\Delta_i$. B$\Delta$I typically reduces data size of SPEC-CPU2006 applications by $\sim50\%$. In these applications, data sections in a cache line are not sorted.

\textbf{CHAIN}. Since both increments and bases of each $k$-mer are sorted and stored consecutively in a DRAM row, we believe they are more compressible than the data of SPEC-CPU2006 applications. Therefore, we propose a novel compression technique, CHAIN, to more aggressively compress an EXMA table. As Figure~\ref{f:dna_chain_all} describes, to compress the increments in a 64B memory line, CHAIN first stores the first increment ($incr_0$). And then, it stores the value difference ($\Delta_i$) between two consecutive increments. So we have $\Delta_i=incr_i-incr_{i-1}$. To decompress an increment $incr_i$ in a 64B memory line, CHAIN simply computes $incr_i=incr_0+\sum_{j=1}^i\Delta_j$. Bases of an EXMA table can be (de)compressed in the same way.

\textbf{Implementation}. The CHAIN compression and decompression require only 64-bit adders. Multiple adders concurrently operate for the CHAIN compression, while the CHAIN decompression requires only one adder for accumulations. The CHAIN decompression slightly prolongs FM-Index search latency but greatly increases FM-Index search throughput. 

\subsubsection{Putting all together}
As Figure~\ref{f:dna_fpga_arch} describes, \ding{182} after receiving FM-Index requests from the CPU, the accelerator stores them in its scheduling queue and performs the first stage scheduling. \ding{183} Based on their $k$-mers, the accelerator checks whether the bases of the requests stored in the queue are in the base cache or not. \ding{184} If misses occur, DRAM accesses are issued to fetch the bases. Otherwise, the accelerator does the second stage scheduling. \ding{185} The accelerator checks whether the MTL index nodes required by the requests in the queue are in the index cache or not, according to both $k$-mer and $pos$ values. \ding{186} If misses happen, DRAM accesses are issued to fetch MTL index nodes. Otherwise, the inference engine computes with MTL index nodes. \ding{187} Until the inference of a leaf node is finished, the accelerator issues a DRAM access to read the increment in the predicted position. If the increment is correct, the computation of $Occ(k\mbox{-}mer,pos)$ is completed. Otherwise, it linearly searches DRAM rows to find the correct increment. \ding{188} All DRAM accesses from the EXMA accelerator are managed by its DMA controller communicating with the CPU MC, which decides whether to pre-charge opened rows based on the requests in the scheduling queue of the accelerator.

\subsubsection{System Integration}
\label{s:integration}

Our EXMA is connected to a CPU processor as a loosely-coupled \textit{non-coherent} accelerator~\cite{Bhardwaj:CAL2019,Grir:IMICRO2018} by a Network-on-Chip (NoC). EXMA accesses DRAM via two DMA-dedicated planes of the NoC, bypassing the cache hierarchy of the CPU. The EXMA data region must be flushed from the CPU cache hierarchy before FM-Index searches start. We chose the non-coherent model~\cite{Bhardwaj:CAL2019} for better performance, since the memory footprint of FM-Index searches is always larger than the CPU LLC capacity.

\begin{table}[t!]
\footnotesize
\setlength{\tabcolsep}{2pt}
\centering
\caption{The hardware configuration of EXMA.}
\vspace{-0.1in}
\begin{tabular}{|c||c|c|c|}
\hline
Component             & Description                & Area ($mm^2$)      & Energy/Op ($pJ$)   \\\hline
Infer. engine         & 4 $8\times8$ PE arrays     & 0.512              & 0.25               \\\hline
Sch. queue            & SRAM, 128-bit$\times$256   & 0.023              & 1.9                \\\hline
Index cache           & SRAM, 32KB, 16-way         & 0.084              & 2.62               \\\hline
Base cache            & eDRAM, 1MB, 8-way          & 0.667              & 17.2               \\\hline
De/compress           & 32 64-bit adders           & 0.091              & 0.21               \\\hline
Sch. \& row           & 2-stage sch. \& dyn. page  & 0.035              & 1.02               \\\hline
DMA ctrl              & adopted from~\cite{Ma:IASIC2009}  & 0.21        & 3.42               \\\hline
\multicolumn{4}{|c|}{EXMA accelerator: area 1.62$mm^2$, and leakage 223.8$mW$}\\\hline\hline
CPU                   & \multicolumn{3}{|l|}{$2.5GHz$, 16-core, $40MB$ LLC, 64 LLC MSHRs}\\\hline\hline
DRAM                  & \multicolumn{3}{|l|}{DDR4-2400, 384GB, 4 channels, 3 DIMMs / channel, }\\
main                  & \multicolumn{3}{|l|}{4 ranks / DIMM, 2 bank groups / rank, 16 chips / rank, }\\
memory							  & \multicolumn{3}{|l|}{2 banks / bank group, FR-FCFS, close-page, row size 2KB}\\	
system                & \multicolumn{3}{|l|}{2 chips / data buffer, $t_{RCD}$-$t_{CAS}$-$t_{RP}$: 16-16-16}\\\hline	
\end{tabular}
\label{t:dna_hardware_overhead}
\vspace{-0.2in}
\end{table}

\subsection{Design Overhead}
The training overhead of a MTL-based index is shown in the Section of \textbf{Training} in \cref{s:mtlindex}. The hardware overhead of the EXMA accelerator is summarized in Table~\ref{t:dna_hardware_overhead}. From~\cite{Gao:ASPLOS2019}, we adopted the inference engine, which runs at $800MHz$ and is synthesized with Synopsys $28nm$ generic library. We quantized the model hierarchy of MTL index with 8-bit without accuracy degradation. A PE has an 8-bit MAC ALU and a 32B register file. The inference engine contains 4 $8\times8$ PE arrays, each of which has a 16KB shared SRAM buffer and an activation unit. The EXMA accelerator also includes a scheduling queue (SRAM CAM) with 512 128-bit entries, a 32KB 16-way SRAM index cache, and a 1MB 8-way eDRAM base cache. We modeled the area and power of memory components including registers, buffers and caches by CACTI~\cite{Jouppi:TVLSIS2014}. We used the same DDR4-2400 DRAM main memory configuration as the recent FM-Index PIM MEDAL~\cite{Huangfu:MICRO2019}. The EXMA accelerator connects to four DRAM channels, with a total 384GB capacity. We adopted DRAMPower~\cite{Chandrasekar:DATE2012} to model the power consumption of our DDR4 main memory.

\section{Experimental Methodology}
\label{s:eandm} 

\textbf{Simulation}. We used gem5-aladdin~\cite{Shao:MICRO2016} to model our CPU baseline and our EXMA accelerator. The configuration of our CPU baseline is shown in Table~\ref{t:dna_hardware_overhead}. We used McPAT~\cite{Li:MICRO2009} to calculate the power consumption of the CPU processor. We implemented the main memory system using DRAMsim2~\cite{Rosenfeld:CAL2011}.

\textbf{Accelerator Baselines}. Besides CPU, we also ran the FM-Index search kernel on an Nvidia Tesla P100 GPU~\cite{N:P100}, a Stratix-V FPGA~\cite{Arram:TCBB2017}, and a $28nm$ ASIC design~\cite{Wang:ICPP2018}. We compared EXMA against recent FM-Index PIM designs, MEDAL~\cite{Huangfu:MICRO2019} and FindeR~\cite{Zokaee:PACT2019}. We used gem5-gpu~\cite{Power:CAL2014} to simulate the GPU, and gem5-aladdin to model the computing units of FPGA, ASIC and PIMs. All their DRAM main memories are implemented by DRAMsim2. The power data of the GPU, FPGA, ASIC and PIMs are adopted from~\cite{N:P100,Arram:TCBB2017,Wang:ICPP2018,Huangfu:MICRO2019,Zokaee:PACT2019}. The power of DRAM main memories is also modeled by DRAMPower.

\textbf{Workloads}. To evaluate EXMA, we adopted FM-Index-based genome analysis applications: BWA-MEM~\cite{Li:BWAMEM2013} for short read alignment, MA~\cite{Schmidt:NC2019} for long read alignment, SGA~\cite{Simpson:GR2012} for read assembly, ExactWordMatch~\cite{Healy:GEN2003} for annotation and a reference-based compression algorithm~\cite{Prochazka:DCC2014}. SGA for long reads uses the FM-Index-based error correcting scheme~\cite{Wang:BMC2018} to reduce errors.

\begin{figure}[t!]
\centering
\includegraphics[width=3.4in]{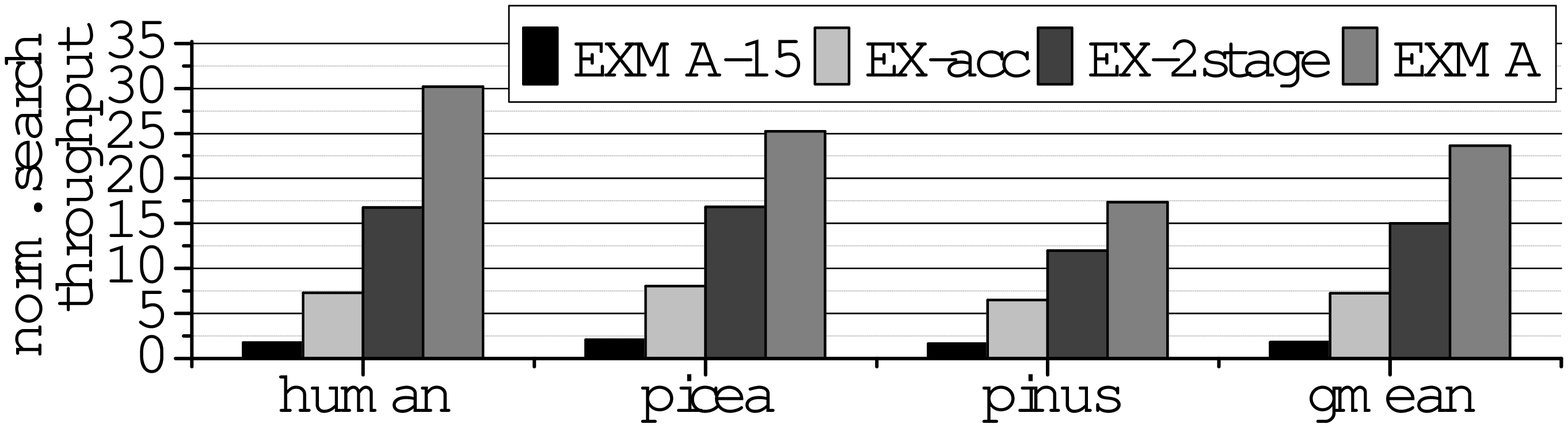}
\vspace{-0.15in}
\caption{The search throughput of EXMA (norm. to \texttt{CPU}).}
\label{f:dna_search_through}
\vspace{-0.25in}
\end{figure}

\textbf{Datasets}. For alignment, annotation and compression, we used human (human, 3G-bp), picea glauca (picea, 20G-bp), and pinus lambertiana (pinus, 31G-bp) genomes as reference genomes. To study short reads, we adopted DWGSim~\cite{Li:BIOFOR2009} to generate 101-bp short reads with $50\times$ coverage. To evaluate long reads, we created long reads (with length of 1K-bp) by PBSIM~\cite{Ono:BIOFOR2012}. The error profiles of reads is summarized in the format of (name, mismatch\%, insertion\%, deletion\%, total error\%), i.e., (Illumina, 0.18\%, 0.01\%, 0.01\%, 0.2\%)~\cite{Schirmer:BMCB2016}, (PacBio, 1.50\%, 9.02\%, 4.49\%, 15.01\%), and (ONT\_2D, 16.50\%, 5.10\%, 8.40\%, 30.0\%)~\cite{Turakhia:ASPLOS2018}.

\textbf{Schemes}. The schemes we studied can be summarized as:
\begin{itemize}[nosep,leftmargin=*]
\item \texttt{CPU}: We ran \texttt{LISA-21} for FM-Index searches in genome applications on our CPU baseline. We also applied B$\Delta$I~\cite{Pekhimenko:PACT2012} compression on LISA data for three datasets.

\item \texttt{EXMA-15}: \texttt{EXMA-15} with the MTL-based index and CHAIN compression is used to replace \texttt{LISA-21} in \texttt{CPU}.

\item \texttt{EX-acc}: We ran \texttt{EXMA-15} on the EXMA accelerator.

\item \texttt{EX-2stage}: 2-stage scheduling is added to \texttt{EX-acc}.

\item \texttt{EXMA}: Dynamic page policy is enabled on \texttt{EX-2stage}.
\end{itemize}

\section{Results and Analysis}
\label{s:reandana}

\textbf{Throughput Comparison against \texttt{CPU}}. As Figure~\ref{f:dna_search_through} shows, we compare FM-Index search throughput of EXMA and \texttt{CPU} by running the seeding of short read alignment, since FM-Index searches consume 99\% of the seeding time in short read alignment. Compared to \texttt{CPU}, on average, \texttt{EXMA-15} improves search throughput by 80\%. Our MTL-based index achieves high accuracy on picea, since the increment distributions of its different $k$-mers are more similar to each other. \texttt{EX-acc} improves search throughput by $7.25\times$ over \texttt{CPU}. Our EXMA accelerator can support more concurrent search operations, while \texttt{CPU} has only a limited number of LLC MSHRs. Compared to \texttt{CPU}, \texttt{EX-2stage} increases search throughput by $15\times$. Pinus with \texttt{EX-2stage} has the smallest throughput improvement. Because the size of the pinus MTL-based index is the largest among 3 datasets, and thus its index cache has the lowest hit rate. On average, \texttt{EXMA} increases search throughput by $23.6\times$ over \texttt{CPU}.

\begin{figure}[t!]
\centering
\includegraphics[width=3.4in]{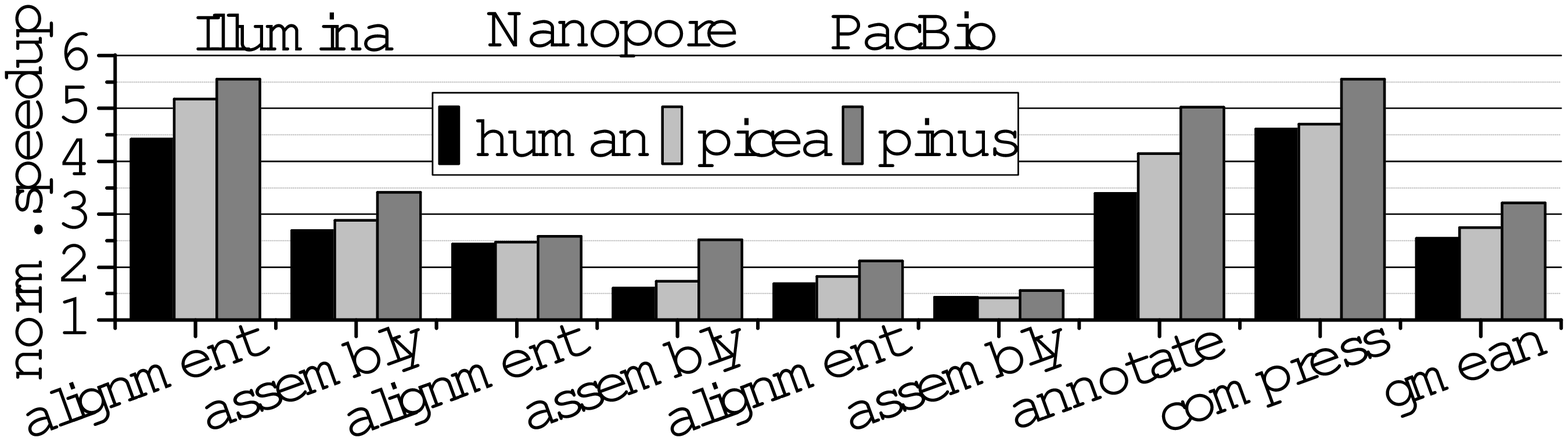}
\vspace{-0.15in}
\caption{The speedup of EXMA in genome analysis (norm. to \texttt{CPU}).}
\label{f:dna_search_application}
\vspace{-0.25in}
\end{figure}

\textbf{Performance Comparison against \texttt{CPU}}. We report and compare the speedup achieved by \texttt{EXMA} in various genome applications in Figure~\ref{f:dna_search_application}, where we list 3 sets of ``alignment and assembly'' for reads generated by Illumina, Nanopore, and PacBio respectively. For each application, although \texttt{EXMA} obtains smaller FM-Index search throughput improvement on larger datasets (Figure~\ref{f:dna_search_through}), e.g., pinus, \texttt{EXMA} improves the application performance more significantly on larger datasets. This is because \texttt{CPU} consumes a larger portion of the execution time of a genome analysis application to perform FM-Index searches when processing larger datasets that introduce more TLB and data cache misses. \texttt{EXMA} achieves larger performance improvement on alignment and assembly for Illumina, annotation, and compression, since FM-Index searches dominate the execution of these applications. On average, \texttt{EXMA} improves the performance of genome applications by $2.5\times\sim 3.2\times$, when processing various datasets.

\begin{figure}[htbp!]
\vspace{-0.15in}
\centering
\includegraphics[width=3.4in]{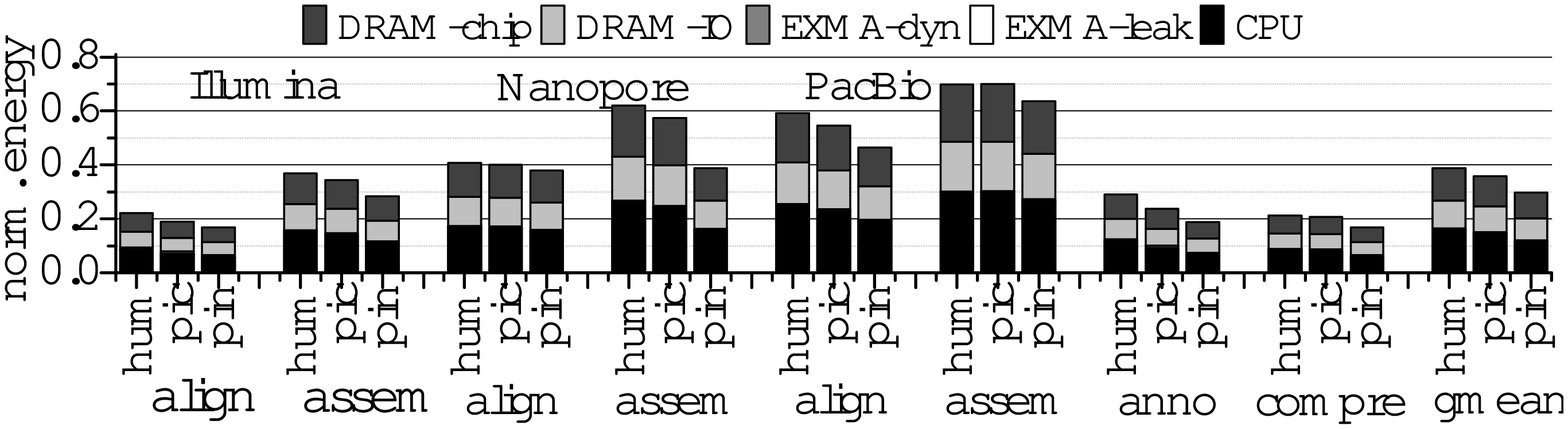}
\vspace{-0.1in}
\caption{The energy reduction of EXMA in genome analysis (norm. to \texttt{CPU}. DRAM-chip/IO indicates the energy of DRAM chips/DDR4 interface. EXMA-leak/dyn is the static/dynamic energy of the EXMA accelerator).}
\label{f:dna_search_energy}
\vspace{-0.1in}
\end{figure}

\textbf{Energy Comparison against \texttt{CPU}}. We compare the energy reduction obtained by \texttt{EXMA} in various genome applications in Figure~\ref{f:dna_search_energy}, where we list 3 sets of ``align(ment) and asse(mbly)'' for reads generated by Illumina, Nanopore, and PacBio respectively. On average, \texttt{EXMA} reduces total energy consumption of genome analysis by $61\%\sim 70\%$ when processing different datasets. The major part of the energy reduction comes from voiding using the CPU processor during FM-Index searches. The more time FM-Index searches consume in a genome analysis application, the more energy reduction \texttt{EXMA} can achieve in that application. On average, the EXMA accelerator consumes only $<3\%$ of the total energy consumption of various genome applications. The vast majority of energy consumption is consumed by the DRAM main memory and the CPU handling non-FM-Index-search parts in genome analysis applications.

\begin{table}[htbp!]
\caption{The comparison of accelerators when processing \texttt{pinus}.}
\label{t:hard_perf_all}
\vspace{-0.1in}
\setlength{\tabcolsep}{1pt}
\scriptsize
\centering
\begin{tabular}{|l|c|c|c|c|c|c|}
\hline
& GPU    & FPGA~\cite{Arram:TCBB2017}      & ASIC~\cite{Wang:ICPP2018}  & MEDAL~\cite{Huangfu:MICRO2019}  & FindeR~\cite{Zokaee:PACT2019} & EXMA    \\ \hline\hline
Algorithm        & LISA-21     & FM-2      & FM-1     & FM-1      & FM-1      & EXMA-15      \\ \hline
Mem (GB)         & 384    & 384    & 384   & 384    & 384    & 384     \\ \hline
Acc Power (W)    & 182    & 11     & 9.4   & 0.011  & 0.28   & 0.89   \\ \hline
Mem Power (W)    & 72     & 72     & 72    & 72     & 72     & 72      \\ \hline
Mbase/s          & 157    & 96     & 34    & 102    & 93     & \textbf{504}    \\ \hline
Mbase/s/Watt     & 0.61   & 0.6    & 0.42  & 1.42   & 1.28   & \textbf{6.9}     \\ \hline
\end{tabular}
\vspace{-0.2in}
\end{table}

\begin{figure*}[t!]
\centering
\begin{minipage}{.25\textwidth}
\centering
\includegraphics[width=1.7in]{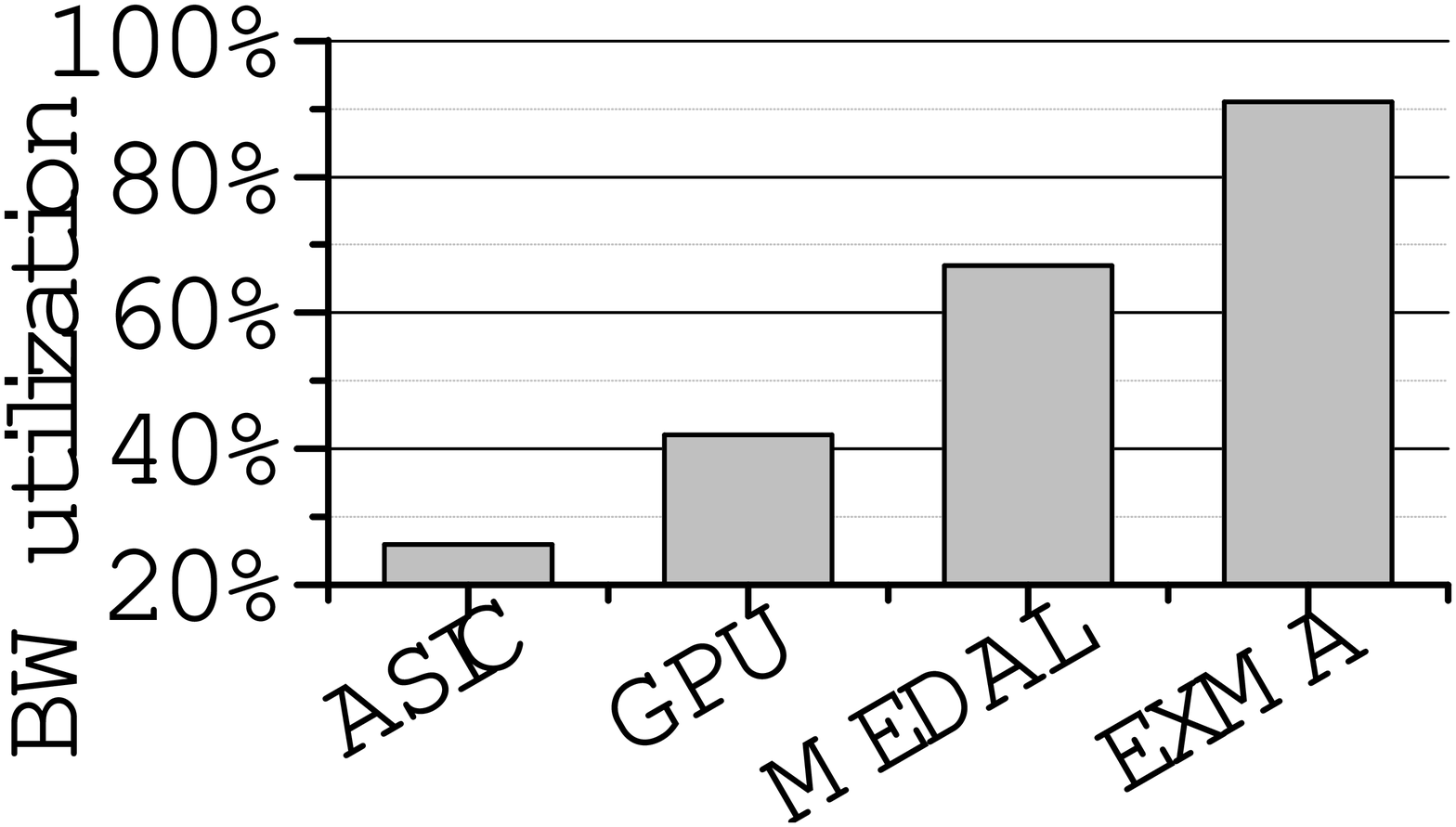}
\vspace{-0.15in}
\caption{Bandwidth utilization}
\label{f:dna_search_bandwidth}
\end{minipage}
\hspace{-0.1in}
\begin{minipage}{0.49\textwidth}
\centering
\includegraphics[width=3.4in]{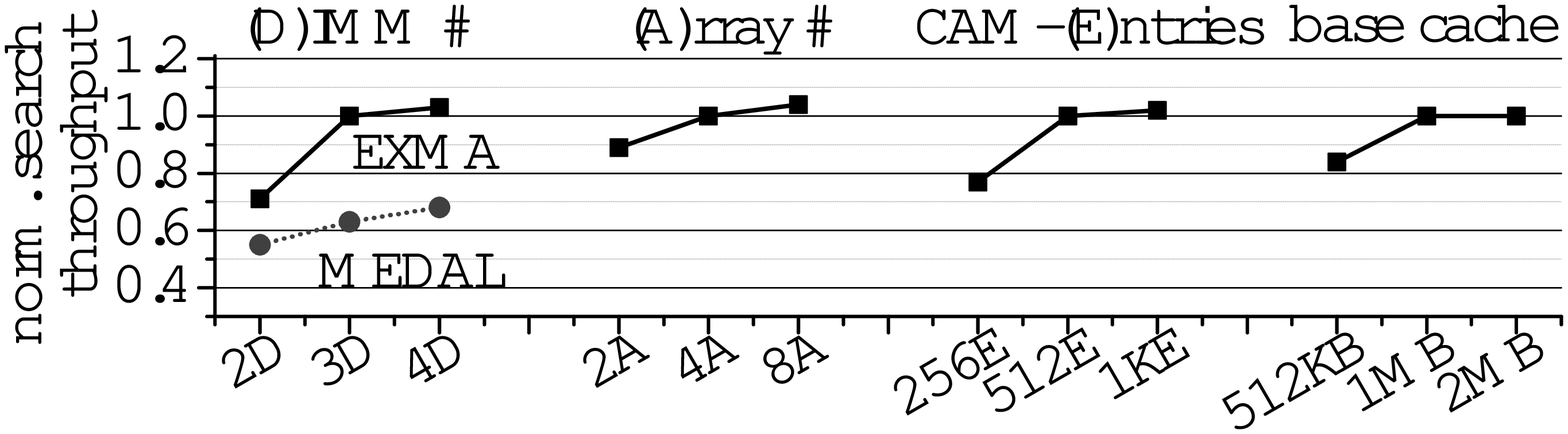}
\vspace{-0.15in}
\caption{Design space exploration (norm. to \texttt{EXMA})}
\label{f:dna_search_study}
\end{minipage}
\hspace{-0.1in}
\begin{minipage}{.25\textwidth}
\centering
\includegraphics[width=1.7in]{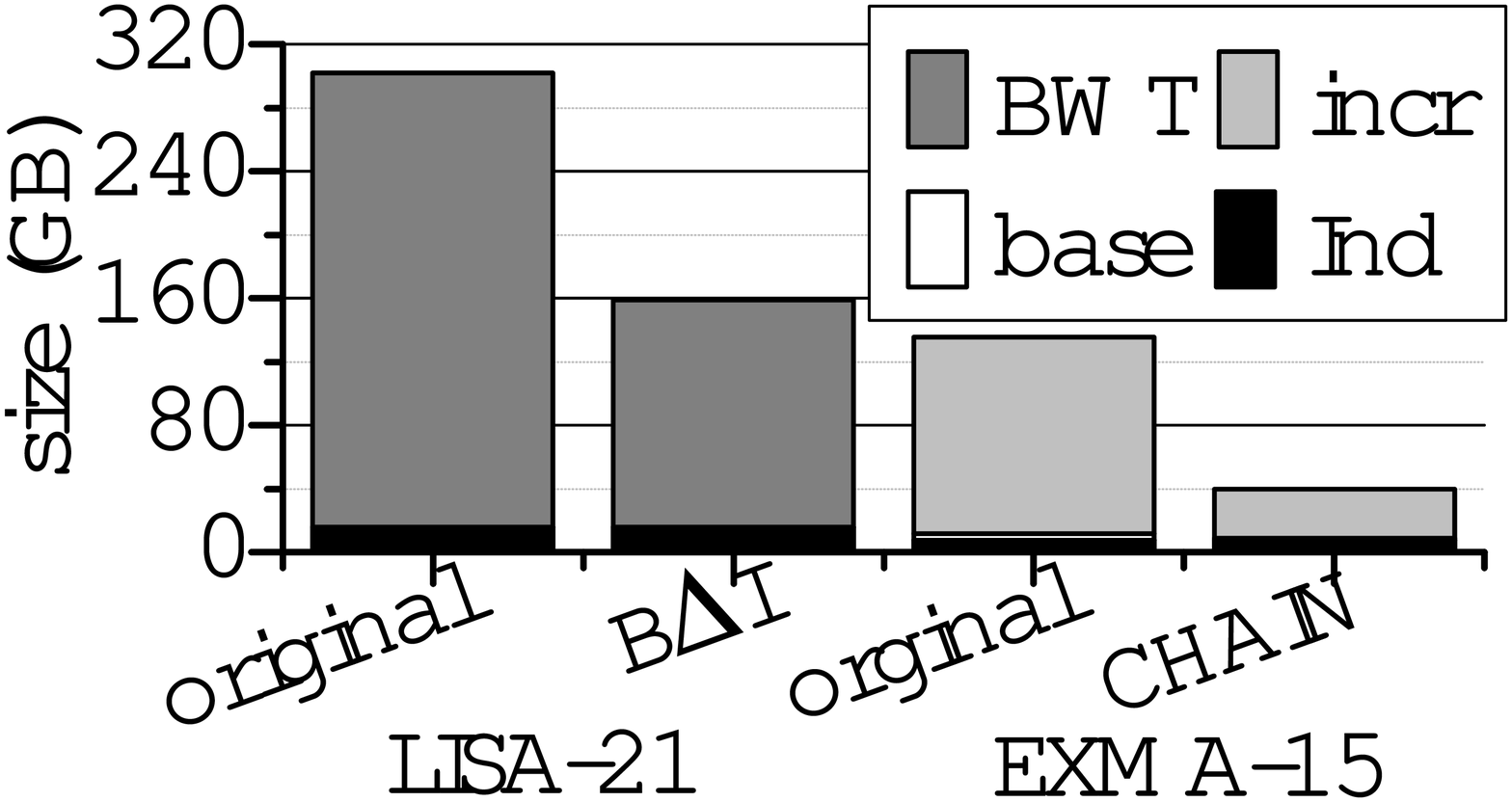}
\vspace{-0.15in}
\caption{CHAIN on \texttt{pinus}}
\label{f:dna_chain_compress}
\end{minipage}
\vspace{-0.25in}
\end{figure*}

\textbf{Comparison against Accelerators}. We evaluated EXMA and compare it against various hardware accelerators including a GPU~\cite{N:P100}, a FPGA~\cite{Arram:TCBB2017}, an ASIC~\cite{Wang:ICPP2018}, and two PIMs~\cite{Huangfu:MICRO2019,Zokaee:PACT2019} when processing \texttt{pinus} in Table~\ref{t:hard_perf_all}. Not all accelerators can run the whole genome applications, so we use ``million base per second'' (Mbase/s) as the performance metric to evaluate only FM-Index search throughput. Different accelerators adopt different search algorithms. We implemented \texttt{LISA-21} on the Tesla P100 GPU. The FPGA design~\cite{Arram:TCBB2017} supports \texttt{FM-2} (conventional 2-step) searches. EXMA performs \texttt{EXMA-15} searches. The other accelerators can conduct only \texttt{FM-1} searches. Since the capacity of internal memories of the GPU (16GB HBM) and the PIM FindeR (2.6GB ReRAM) is smaller than the \texttt{pinus} data size. We provide the same DDR4 main memory configuration shown in Table~\ref{t:dna_hardware_overhead} to all accelerators. The power values of both the accelerator (Acc Power) and the DDR4 main memory (Mem Power) are listed in Table~\ref{t:hard_perf_all}. The search performance is decided by two factors, i.e., the memory bandwidth utilization, and the number of DNA symbols searched during each iteration. For the bandwidth utilization, only EXMA supports dynamic page policy, while the other use only close page policy. MEDAL can support chip-level parallelism, but its search throughput is limited by the address bus. So EXMA achieves the highest bandwidth utilization. For the number of DNA symbols searched during each iteration, only the GPU and EXMA can process $>2$ DNA symbols in each iteration. But the low accuracy of learned index makes the GPU to search many unnecessary IP-BWT entries to find the correct one. Therefore, EXMA obtains the best search throughput. The throughput per Watt of the GPU and FPGA designs is low, since the power consumption of their computing parts is not trivial. For the PIMs and EXMA, the power consumption is dominated by the DRAM main memory. Compared to the PIM MEDAL, EXMA improves search throughput per Watt by $4.8\times$.

\textbf{Bandwidth utilization}. Figure~\ref{f:dna_search_bandwidth} shows the comparison of bandwidth utilization, which is defined as the ratio between the data capacity fetched from DRAM and total DRAM bandwidth. ASIC using \texttt{FM-1} has only 26\% of the total DRAM bandwidth, since it uses close-page policy and fetches only 64B after activating a 2KB row. GPU implementing \texttt{LISA-21} fetches entire rows from host memory, so it achieves higher bandwidth utilization. MEDAL increases bandwidth utilization by activating each individual chips. However, due to conflicts on the address bus, MEDAL uses only 67\% of the DRAM bandwidth. In contrast, EXMA obtains 91\% bandwidth utilization by switching between close-page and open-page policies.

\textbf{DIMM number}. We studied the sensitivity of EXMA and MEDAL to the DIMM number in Figure~\ref{f:dna_search_study}. With 2 DIMMs, EXMA improves search throughput by 29\% over MEDAL. By having 3 DIMMs, EXMA linearly scales its throughput up by 40\%, since a single EXMA accelerator can maintain all the DIMMs. However, MEDAL increases its throughput by only 14.5\% with 3 DIMMs. Each MEDAL PIM accelerator sits on a DIMM. More DIMMs bring more ranks. MEDAL suffers from non-trivial inter-DIMM communication overhead. When the number of DIMM increases to 4, the search throughput of neither EXMA nor MEDAL increases significantly. The data bus (bandwidth utilization) of EXMA is saturated, while the address bus of MEDAL is saturated.

\textbf{PE Array number}. We show search throughput of EXMA with a varying number of PE arrays in Figure~\ref{f:dna_search_study}. Two PE arrays of EXMA already achieve 89\% of the search throughput of the configuration with four PE arrays. This is because the computations of MTL-based indexes are not intensive. Further increasing the PE array number to 8 increases search throughput by only 3\% over the configuration with four PE arrays. So we selected 4 PE arrays in our baseline configuration.

\textbf{CAM \& Cache}. We explored search throughput of EXMA with varying CAM and base cache sizes in Figure~\ref{f:dna_search_study}. We use a CAM consisting of 256, 512, and 1024 entries to serve as the scheduling queue of EXMA. A 256-entry CAM cannot fully satisfy 2-stage scheduling and scheduling for dynamic page policy, and thus achieves only 77\% of search throughput of the configuration with a 512-entry CAM. Further increasing the CAM entry number to 1K improves search throughput by only 2\% over the configuration with a 512-entry CAM. Compared to the index cache, the search throughput is more sensitive to the capacity of the base cache, since the global buffer and register file of PE arrays can temporarily store MTL-based index nodes. We tried 512KB, 1MB and 2MB for the base cache. The search throughput stops increasing significantly, when the base cache capacity reaches 1MB. So we selected a 512-entry CAM and a 1MB base cache in our baseline configuration.

\textbf{CHAIN}. We show the compression result of CHAIN on \texttt{pinus} in Figure~\ref{f:dna_chain_compress}. Since the size of the IP-BWT table of \texttt{LISA-21} is proportional to its step number, the total data size of \texttt{LISA-21} (original) is $2.2\times$ larger than that of \texttt{EXMA-15} (original). After B$\Delta$I compresses the data size of \texttt{LISA-21} to 50\%, i.e., 152GB. On the contrary, CHAIN compresses the data size of \texttt{EXMA-15} to only 25\%, i.e., 40GB. We observed similar compression rates of B$\Delta$I and CHAIN on the other genome datasets.

\section{Conclusion}
\label{s:con}
Though state-of-the-art genome analysis adopts FM-Index to process exact-matches, FM-Index is notorious of random memory access patterns. In this paper, we first present a row-buffer-friendly and highly-compressible EXMA table with a MTL-based index to process multiple DNA symbols by activating a DRAM row during each search iteration. And then, we build a hardware accelerator to process FM-Index searches on a EXMA table. Compared to the state-of-the-art FM-Index PIM MEDAL, EXMA improves search throughput by $4.9\times$, and enhances search throughput per Watt by $4.8\times$.

\bibliographystyle{IEEEtran}
\bibliography{genomics}

\begin{thebibliography}{10}
\providecommand{\url}[1]{#1}
\csname url@samestyle\endcsname
\providecommand{\newblock}{\relax}
\providecommand{\bibinfo}[2]{#2}
\providecommand{\BIBentrySTDinterwordspacing}{\spaceskip=0pt\relax}
\providecommand{\BIBentryALTinterwordstretchfactor}{4}
\providecommand{\BIBentryALTinterwordspacing}{\spaceskip=\fontdimen2\font plus
\BIBentryALTinterwordstretchfactor\fontdimen3\font minus
  \fontdimen4\font\relax}
\providecommand{\BIBforeignlanguage}[2]{{%
\expandafter\ifx\csname l@#1\endcsname\relax
\typeout{** WARNING: IEEEtran.bst: No hyphenation pattern has been}%
\typeout{** loaded for the language `#1'. Using the pattern for}%
\typeout{** the default language instead.}%
\else
\language=\csname l@#1\endcsname
\fi
#2}}
\providecommand{\BIBdecl}{\relax}
\BIBdecl

\bibitem{Schirmer:NAR2015}
M.~Schirmer, U.~Z. Ijaz, R.~D'Amore, N.~Hall, W.~T. Sloan, and C.~Quince,
  ``Insight into biases and sequencing errors for amplicon sequencing with the
  illumina miseq platform,'' \emph{Nucleic acids research}, vol.~43, no.~6, pp.
  e37--e37, 2015.

\bibitem{Mosher:JMM2014}
J.~J. Mosher, B.~Bowman, E.~L. Bernberg, O.~Shevchenko, J.~Kan, J.~Korlach, and
  L.~A. Kaplan, ``Improved performance of the pacbio smrt technology for 16s
  rdna sequencing,'' \emph{Journal of microbiological methods}, vol. 104, pp.
  59--60, 2014.

\bibitem{Eisenstein:Oxford2012}
M.~Eisenstein, ``Oxford nanopore announcement sets sequencing sector abuzz,''
  \emph{Nature Biotechnology}, vol.~30, no.~4, pp. 295--297, 2012.

\bibitem{Merker:Nature2018}
J.~D. Merker, A.~M. Wenger, T.~Sneddon, M.~Grove, Z.~Zappala, L.~Fresard,
  D.~Waggott, S.~Utiramerur, Y.~Hou, K.~S. Smith \emph{et~al.}, ``Long-read
  genome sequencing identifies causal structural variation in a mendelian
  disease,'' \emph{Genetics in Medicine}, vol.~20, no.~1, p. 159, 2018.

\bibitem{Ma:TBIO2017}
X.~Ma, M.~Mau, and T.~F. Sharbel, ``Genome editing for global food security,''
  \emph{Trends in biotechnology}, 2017.

\bibitem{Hoenen:EID2016}
T.~Hoenen, A.~Groseth, K.~Rosenke, R.~J. Fischer, A.~Hoenen, S.~D. Judson,
  C.~Martellaro, D.~Falzarano, A.~Marzi, R.~B. Squires \emph{et~al.},
  ``Nanopore sequencing as a rapidly deployable ebola outbreak tool,''
  \emph{Emerging infectious diseases}, vol.~22, no.~2, p. 331, 2016.

\bibitem{Quick:NP2017}
J.~Quick, N.~D. Grubaugh, S.~T. Pullan, I.~M. Claro, A.~D. Smith,
  K.~Gangavarapu, G.~Oliveira, R.~Robles-Sikisaka, T.~F. Rogers, N.~A. Beutler
  \emph{et~al.}, ``Multiplex pcr method for minion and illumina sequencing of
  zika and other virus genomes directly from clinical samples,'' \emph{nature
  protocols}, vol.~12, no.~6, p. 1261, 2017.

\bibitem{Zhu:NEJM2020}
N.~Zhu, D.~Zhang, W.~Wang, X.~Li, B.~Yang, J.~Song, X.~Zhao, B.~Huang, W.~Shi,
  R.~Lu \emph{et~al.}, ``A novel coronavirus from patients with pneumonia in
  china, 2019,'' \emph{New England Journal of Medicine}, 2020.

\bibitem{Canzar:IEEE2017}
S.~Canzar and S.~L. Salzberg, ``Short read mapping: An algorithmic tour,''
  \emph{Proceedings of the IEEE}, vol. 105, no.~3, pp. 436--458, 2017.

\bibitem{Turakhia:ASPLOS2018}
Y.~Turakhia, G.~Bejerano, and W.~J. Dally, ``Darwin: A genomics co-processor
  provides up to 15,000x acceleration on long read assembly,'' in \emph{ACM
  Architectural Support for Programming Languages and Operating Systems}, 2018.

\bibitem{Fuijiki:ISCA2018}
D.~Fuijiki, A.~Subramaniyan, T.~Zhang, Y.~Zheng, R.~Das, D.~Blaauw, and
  S.~Narayanasamy, ``Genax: A genome sequencing accelerator,'' in
  \emph{IEEE/ACM International Symposium on Computer Architecture}, 2018.

\bibitem{Li:BWAMEM2013}
H.~Li, ``Aligning sequence reads, clone sequences and assembly contigs with
  bwa-mem,'' \emph{arXiv preprint arXiv:1303.3997}, 2013.

\bibitem{Chang:FCCM2016}
M.~C.~F. Chang, Y.~T. Chen, J.~Cong, P.~T. Huang, C.~L. Kuo, and C.~H. Yu,
  ``The smem seeding acceleration for dna sequence alignment,'' in \emph{IEEE
  International Symposium on Field-Programmable Custom Computing Machines
  FCCM}, 2016, pp. 32--39.

\bibitem{Zokaee:PACT2019}
F.~Zokaee, M.~Zhang, and L.~Jiang, ``Finder: Accelerating fm-index-based exact
  pattern matching in genomic sequences through reram technology,'' in
  \emph{International Conference on Parallel Architectures and Compilation
  Techniques}, 2019, pp. 284--295.

\bibitem{Huangfu:MICRO2019}
W.~Huangfu, X.~Li, S.~Li, X.~Hu, P.~Gu, and Y.~Xie, ``Medal: Scalable dimm
  based near data processing accelerator for dna seeding algorithm,'' in
  \emph{IEEE/ACM International Symposium on Microarchitecture}, 2019, p.
  587–599.

\bibitem{Schmidt:NC2019}
M.~Schmidt, K.~Heese, and A.~Kutzner, ``Accurate high throughput alignment via
  line sweep-based seed processing,'' \emph{Nature communications}, vol.~10,
  no.~1, pp. 1--10, 2019.

\bibitem{Nag:MICRO2019}
A.~Nag, C.~Ramachandra, R.~Balasubramonian, R.~Stutsman, E.~Giacomin,
  H.~Kambalasubramanyam, and P.-E. Gaillardon, ``Gencache: Leveraging in-cache
  operators for efficient sequence alignment,'' in \emph{IEEE/ACM International
  Symposium on Microarchitecture}, 2019, pp. 334--346.

\bibitem{Kaplan:MICRO2017}
R.~Kaplan, L.~Yavits, R.~Ginosar, and U.~Weiser, ``A resistive cam
  processing-in-storage architecture for dna sequence alignment,'' \emph{IEEE
  Micro}, 2017.

\bibitem{Madhavan:ISCA2014}
A.~Madhavan, T.~Sherwood, and D.~Strukov, ``Race logic: A hardware acceleration
  for dynamic programming algorithms,'' in \emph{IEEE/ACM International
  Symposium on Computer Architecture}, 2014.

\bibitem{Enzo:ICBB2017}
E.~Rucci, C.~Garcia, G.~Botella, A.~De~Giusti, M.~Naiouf, and M.~Prieto-Matias,
  ``Accelerating smith-waterman alignment of long dna sequences with opencl on
  fpga,'' in \emph{International Conference on Bioinformatics and Biomedical
  Engineering}.\hskip 1em plus 0.5em minus 0.4em\relax Springer, 2017, pp.
  500--511.

\bibitem{Luo:PLOS2013}
R.~Luo, T.~Wong, J.~Zhu, C.-M. Liu, X.~Zhu, E.~Wu, L.-K. Lee, H.~Lin, W.~Zhu,
  D.~W. Cheung \emph{et~al.}, ``Soap3-dp: fast, accurate and sensitive
  gpu-based short read aligner,'' \emph{PloS one}, vol.~8, no.~5, p. e65632,
  2013.

\bibitem{Burrows:HSRR1994}
M.~Burrows and D.~J. Wheeler, ``A block-sorting lossless data compression
  algorithm,'' 1994.

\bibitem{Ahmed:ICBB2016}
N.~Ahmed, K.~Bertels, and Z.~Al-Ars, ``A comparison of seed-and-extend
  techniques in modern dna read alignment algorithms,'' in \emph{IEEE
  International Conference on Bioinformatics and Biomedicine}, 2016, pp.
  1421--1428.

\bibitem{Simpson:GR2012}
J.~T. Simpson and R.~Durbin, ``Efficient de novo assembly of large genomes
  using compressed data structures,'' \emph{Genome research}, vol.~22, no.~3,
  2012.

\bibitem{Healy:GEN2003}
J.~Healy, E.~E. Thomas, J.~T. Schwartz, and M.~Wigler, ``Annotating large
  genomes with exact word matches,'' \emph{Genome research}, vol.~13, no.~10,
  pp. 2306--2315, 2003.

\bibitem{Prochazka:DCC2014}
P.~Prochazka and J.~Holub, ``Compressing similar biological sequences using
  fm-index,'' in \emph{Data Compression Conference}, 2014, pp. 312--321.

\bibitem{Chacon:TCBB2015}
A.~{Chacon}, S.~{Marco-Sola}, A.~{Espinosa}, P.~{Ribeca}, and J.~C. {Moure},
  ``Boosting the fm-index on the gpu: Effective techniques to mitigate random
  memory access,'' \emph{IEEE/ACM Transactions on Computational Biology and
  Bioinformatics}, vol.~12, no.~5, pp. 1048--1059, 2015.

\bibitem{Ho:WSMN2019}
D.~Ho, J.~Ding, S.~Misra, N.~Tatbul, V.~Nathan, V.~Md, and T.~Kraska, ``Lisa:
  Towards learned dna sequence search,'' in \emph{Workshop on Systems for ML at
  NeurIPS 2019}, 2019.

\bibitem{Kraska:ICMD2018}
T.~Kraska, A.~Beutel, E.~H. Chi, J.~Dean, and N.~Polyzotis, ``The case for
  learned index structures,'' in \emph{ACM International Conference on
  Management of Data}, 2018, p. 489–504.

\bibitem{Arram:TCBB2017}
J.~Arram, T.~Kaplan, W.~Luk, and P.~Jiang, ``Leveraging fpgas for accelerating
  short read alignment,'' \emph{IEEE/ACM Transactions on Computational Biology
  and Bioinformatics}, 2017.

\bibitem{Wu:ITBCS2017}
Y.~C. Wu, C.~H. Chang, J.~H. Hung, and C.~H. Yang, ``A 135-mw fully integrated
  data processor for next-generation sequencing,'' \emph{IEEE Transactions on
  Biomedical Circuits and Systems}, 2017.

\bibitem{Quail:BMC2012}
M.~A. Quail, M.~Smith, P.~Coupland, T.~D. Otto, S.~R. Harris, T.~R. Connor,
  A.~Bertoni, H.~P. Swerdlow, and Y.~Gu, ``A tale of three next generation
  sequencing platforms: comparison of ion torrent, pacific biosciences and
  illumina miseq sequencers,'' \emph{BMC genomics}, vol.~13, no.~1, p. 341,
  2012.

\bibitem{Wang:BMC2018}
J.~R. Wang, J.~Holt, L.~McMillan, and C.~D. Jones, ``Fmlrc: Hybrid long read
  error correction using an fm-index,'' \emph{BMC bioinformatics}, vol.~19,
  no.~1, p.~50, 2018.

\bibitem{Huang:BMC2017}
Y.-T. Huang and Y.-W. Huang, ``An efficient error correction algorithm using
  fm-index,'' \emph{BMC bioinformatics}, vol.~18, no.~1, p. 524, 2017.

\bibitem{Li:BIOINFO2012}
H.~Li, ``Exploring single-sample snp and indel calling with whole-genome de
  novo assembly,'' \emph{Bioinformatics}, vol.~28, no.~14, pp. 1838--1844,
  2012.

\bibitem{Chacon:PCS2013}
A.~Chac{\'o}n, J.~C. Moure, A.~Espinosa, and P.~Hern{\'a}ndez, ``n-step
  fm-index for faster pattern matching,'' \emph{Procedia Computer Science},
  vol.~18, pp. 70--79, 2013.

\bibitem{Wang:ICPP2018}
Y.~Wang, X.~Li, D.~Zang, G.~Tan, and N.~Sun, ``Accelerating fm-index search for
  genomic data processing,'' in \emph{International Conference on Parallel
  Processing}, 2018.

\bibitem{Shaahin:DAC2019}
S.~Angizi, J.~Sun, W.~Zhang, and D.~Fan, ``Aligns: A processing-in-memory
  accelerator for dna short read alignment leveraging sot-mram,'' in
  \emph{ACM/IEEE Design Automation Conference}, 2019, pp. 1--6.

\bibitem{Angizi:DATE2020}
S.~{Angizi}, J.~{Sun}, W.~{Zhang}, and D.~{Fan}, ``Pim-aligner: A
  processing-in-mram platform for biological sequence alignment,'' in
  \emph{Design, Automation Test in Europe Conference \& Exhibition}, 2020, pp.
  1265--1270.

\bibitem{JEDEC:JESD79-4C}
\BIBentryALTinterwordspacing
JEDEC, ``Ddr4 sdram standard jesd79-4c,'' 2020. [Online]. Available:
  \url{https://www.jedec.org/standards-documents/docs/jesd79-4a}
\BIBentrySTDinterwordspacing

\bibitem{Ruder:ARXIV2017}
S.~Ruder, ``An overview of multi-task learning in deep neural networks,''
  \emph{CoRR}, vol. abs/1706.05098, 2017.

\bibitem{Kendall:CVPR2018}
A.~Kendall, Y.~Gal, and R.~Cipolla, ``Multi-task learning using uncertainty to
  weigh losses for scene geometry and semantics,'' in \emph{IEEE conference on
  computer vision and pattern recognition}, 2018, pp. 7482--7491.

\bibitem{Bilen:NIPS2016}
H.~Bilen and A.~Vedaldi, ``Integrated perception with recurrent multi-task
  neural networks,'' in \emph{Advances in neural information processing
  systems}, 2016, pp. 235--243.

\bibitem{Stein:Stan1956}
C.~Stein, ``Inadmissibility of the usual estimator for the mean of a
  multivariate normal distribution,'' Stanford University, Tech. Rep., 1956.

\bibitem{Kutner:ALSM2005}
M.~H. Kutner, C.~J. Nachtsheim, J.~Neter, W.~Li \emph{et~al.}, \emph{Applied
  linear statistical models}.\hskip 1em plus 0.5em minus 0.4em\relax
  McGraw-Hill Irwin New York, 2005, vol.~5.

\bibitem{Argyriou:NIPS2007}
A.~Argyriou, T.~Evgeniou, and M.~Pontil, ``Multi-task feature learning,'' in
  \emph{Advances in neural information processing systems}, 2007, pp. 41--48.

\bibitem{Gao:ASPLOS2019}
M.~Gao, X.~Yang, J.~Pu, M.~Horowitz, and C.~Kozyrakis, ``Tangram: Optimized
  coarse-grained dataflow for scalable nn accelerators,'' in \emph{ACM
  International Conference on Architectural Support for Programming Languages
  and Operating Systems}, 2019, pp. 807--820.

\bibitem{Okabayashi:VLSI1990}
I.~{Okabayashi}, H.~{Kotani}, and H.~{Kadota}, ``A proposed structure of a 4
  mbit content-addressable and sorting memory,'' in \emph{Symposium on VLSI
  Circuits Digest of Technical Papers}, 1990, pp. 109--110.

\bibitem{Pekhimenko:PACT2012}
G.~Pekhimenko, V.~Seshadri, O.~Mutlu, M.~A. Kozuch, P.~B. Gibbons, and T.~C.
  Mowry, ``Base-delta-immediate compression: Practical data compression for
  on-chip caches,'' in \emph{IEEE International Conference on Parallel
  Architectures and Compilation Techniques}, 2012.

\bibitem{Bhardwaj:CAL2019}
K.~{Bhardwaj}, M.~{Havasi}, Y.~{Yao}, D.~M. {Brooks}, J.~M.~H. {Lobato}, and
  G.~{Wei}, ``Determining optimal coherency interface for many-accelerator socs
  using bayesian optimization,'' \emph{IEEE Computer Architecture Letters},
  vol.~18, no.~2, pp. 119--123, 2019.

\bibitem{Grir:IMICRO2018}
D.~Giri, P.~Mantovani, and L.~P. Carloni, ``Accelerators and coherence: An soc
  perspective,'' \emph{IEEE Micro}, vol.~38, no.~6, pp. 36--45, 2018.

\bibitem{Ma:IASIC2009}
G.~Ma and H.~He, ``Design and implementation of an advanced dma controller on
  amba-based soc,'' in \emph{IEEE International Conference on ASIC}, 2009, pp.
  419--422.

\bibitem{Jouppi:TVLSIS2014}
N.~P. Jouppi, A.~B. Kahng, N.~Muralimanohar, and V.~Srinivas, ``Cacti-io: Cacti
  with off-chip power-area-timing models,'' \emph{IEEE Transactions on Very
  Large Scale Integration Systems}, vol.~23, no.~7, pp. 1254--1267, 2014.

\bibitem{Chandrasekar:DATE2012}
\BIBentryALTinterwordspacing
K.~Chandrasekar, C.~Weis, Y.~Li, B.~Akesson, N.~Wehn, and K.~Goossens,
  ``Drampower: Open-source dram power \& energy estimation tool,'' 2012.
  [Online]. Available: \url{https://github.com/tukl-msd/DRAMPower}
\BIBentrySTDinterwordspacing

\bibitem{Shao:MICRO2016}
Y.~S. Shao, S.~L. Xi, V.~Srinivasan, G.-Y. Wei, and D.~Brooks, ``Co-designing
  accelerators and soc interfaces using gem5-aladdin,'' in \emph{IEEE/ACM
  International Symposium on Microarchitecture (MICRO)}, 2016, pp. 1--12.

\bibitem{Li:MICRO2009}
S.~Li, J.~H. Ahn, R.~D. Strong, J.~B. Brockman, D.~M. Tullsen, and N.~P.
  Jouppi, ``Mcpat: an integrated power, area, and timing modeling framework for
  multicore and manycore architectures,'' in \emph{IEEE/ACM International
  Symposium on Microarchitecture}, 2009, pp. 469--480.

\bibitem{Rosenfeld:CAL2011}
P.~Rosenfeld, E.~Cooper-Balis, and B.~Jacob, ``Dramsim2: A cycle accurate
  memory system simulator,'' \emph{IEEE computer architecture letters},
  vol.~10, no.~1, pp. 16--19, 2011.

\bibitem{N:P100}
\BIBentryALTinterwordspacing
NVIDIA, ``Nvidia tesla p100,'' 2016. [Online]. Available:
  \url{https://images.nvidia.com/content/pdf/tesla/whitepaper/pascal-architecture-whitepaper.pdf}
\BIBentrySTDinterwordspacing

\bibitem{Power:CAL2014}
J.~Power, J.~Hestness, M.~Orr, M.~Hill, and D.~Wood, ``gem5-gpu: A
  heterogeneous cpu-gpu simulator,'' \emph{Computer Architecture Letters},
  vol.~13, no.~1, Jan 2014.

\bibitem{Li:BIOFOR2009}
H.~Li, B.~Handsaker, A.~Wysoker, T.~Fennell, J.~Ruan, N.~Homer, G.~Marth,
  G.~Abecasis, and R.~Durbin, ``The sequence alignment/map format and
  samtools,'' \emph{Bioinformatics}, vol.~25, no.~16, pp. 2078--2079, 2009.

\bibitem{Ono:BIOFOR2012}
Y.~Ono, K.~Asai, and M.~Hamada, ``Pbsim: Pacbio reads simulator—toward
  accurate genome assembly,'' \emph{Bioinformatics}, vol.~29, no.~1, pp.
  119--121, 2012.

\bibitem{Schirmer:BMCB2016}
M.~Schirmer, R.~D'Amore, U.~Z. Ijaz, N.~Hall, and C.~Quince, ``Illumina error
  profiles: resolving fine-scale variation in metagenomic sequencing data,''
  \emph{BMC bioinformatics}, vol.~17, no.~1, p. 125, 2016.

\end{thebibliography}

\end{document}